\documentclass[a4paper,11pt]{article}
\pdfoutput=1 

\usepackage{jheppub} 

\usepackage[T1]{fontenc} 
\usepackage{mathtools}
\usepackage{amsmath}
\usepackage{amssymb}
\usepackage{wasysym}
\usepackage{amsfonts}
\usepackage{multirow}
\usepackage{tikz}
\usetikzlibrary{arrows,quotes,angles}
\usepackage{pgfplots}
\usepackage[colorinlistoftodos]{todonotes}
\pgfplotsset{width=10cm,compat=1.9}
\usepackage{physics}
\usepackage{caption}
\usepackage{yfonts}
\usepackage{euscript}
\usepackage{subcaption}
\usepackage{tablefootnote}
\usepackage{yfonts}
\usepackage{graphicx}
\usepackage{adjustbox}
\usepackage{booktabs}
\usepackage{hyperref}
\usepackage{cleveref}
\usepackage{float}
\Crefname{equation}{Eq.}{Eqs.}
\Crefname{figure}{Fig.}{Figs.}
\Crefname{definition}{Def.}{Defs.}

\newtheorem{definition}{Definition}

\usepackage[normalem]{ulem}

\title{\boldmath Canonical General Relativity and Emergent Geometry}

\author[a,1]{Philip Tee\note{Corresponding author.}}

\emailAdd{ptee2@asu.edu}

\affiliation[a]{Beyond Center, Arizona State University, \\ Tempe, Arizona 85287}

\abstract{Ising models of emergent geometry are well known to possess ground states with many of the desired features of a low dimensional, Ricci flat vacuum.
Further, excitations of these ground states can be shown to replicate the  quantum dynamics of a free particle in the continuum limit.
It would be a significant next step in the development of emergent Ising models to link them to an underlying physical theory that has General Relativity as its continuum limit.
In this work we investigate how the canonical formulation of General Relativity can be used to construct such a discrete Hamiltonian using recent results in discrete differential geometry.
We are able to demonstrate that the Ising models of emergent geometry are closely related to the model we propose, which we term the  Canonical Ising Model, and may be interpreted as an approximation of discretized canonical general relativity.}

\begin{document} 
\maketitle
\flushbottom

\section{Introduction}\label{sec:introduction}
\subsection{Background}

Finding a complete and consistent theory that reconciles Quantum Mechanics (QM) and General Relativity (GR) has been an elusive goal since the very first attempts \cite{bronstein1936kvantovanie} made during the `miracle decades' of theoretical physics at the start of the last century.
Independently both theories have been remarkably successful, but to date no consistent theory exists that successfully unites both \cite{graf2018hawking}.
At the heart of the problem is a profound inconsistency between how space and time are treated.
For QM and Quantum Field Theory (QFT) space and time are transcendent quantities, input by hand into the theories, whereas GR {\sl is essentially} a theory of the geometry of  spacetime.

Recently new theoretical models that capitalize on advances in network science have emerged that attempt to reconcile GR and QM using an emergent geometry.
These `Ising Geometry' (IG) models were originally proposed by Trugenberger \cite{trugenberger2015quantum}, subsequently extended and applied to quantum gravity  \cite{trugenberger2016random,trugenberger2017combinatorial,kelly2019self,tee2020dynamics,tee2021quantum}, and are central to the  Combinatorial Quantum Gravity (CQG) program.
A fundamental precept of these models is a discrete quantum structure, implicitly or explicitly assumed to be at the Planck length scale.
This approach can be viewed as building upon other theories that imply a discrete structure to spacetime, including Quantum Graphity \cite{konopka2006quantum,konopka2008quantum}, Causal Set theory \cite{bombelli1987space,dowker2006causal}, and Loop Quantum Gravity \cite{rovelli2014covariant,smolin2004invitation,smolin2011unimodular}.
Although not normally associated with discrete geometries, type IIB String theories also admit a  `matrix model' formulation \cite{ishibashi1997large,klinkhamer2020iib}, which proposes a Lagrangian with an $SU(N)$ gauge symmetry where $N$ is very large.
In matrix models, conventional spacetime `emerges', with Lorentzian signature metrics, when one identifies  the eigenvalues of the $N \times N$ gauge field matrices with spacetime locations, with the limit of $N \rightarrow \infty$, recovering a smooth manifold.
Cao {\sl et al} \cite{cao2017space} pointed out that it is also possible to cause geometry to emerge from the quantum entanglement of a large number of interacting Hilbert spaces, yielding a discrete graph structure for the geometry.
Finally, and more recently, Wolfram {\sl et al} proposed an entirely abstract method of emerging a discrete spacetime based upon `branchial graphs' \cite{gorard2020some,gorard2020zx}, although entirely different from the models described here they share the commonality of representing the emerged geometry as a graph.

In this work we will focus upon the IG models so named because they propose that spacetime emerges from a disordered and  independent collection of quantum information bits subject to a ferromagnetic and Ising like interaction.
For the precise details we refer the reader to the cited literature \cite{trugenberger2015quantum,trugenberger2016random,tee2020dynamics}, but in principle the interaction between information bits creates the fabric of a geometry expressed as a graph, with entangled spins sharing an edge.
To prevent a fully connected graph emerging there are  anti-ferromagnetic terms introduced into the  Hamiltonians of these models that disfavor edges, and higher order structures such as triangles, that would impair locality.
The models possess many attractive features, including a regular and highly local ground state, the emergence of a low preferential dimension as evidenced by a convergence of extrinsic and intrinsic dimensions, and the capacity to support the modelling of matter by the presence of stable defects in the mesh \cite{tee2020dynamics}.

These models  do not yet amount to a complete and consistent theory of spacetime, not least because there is no accepted and established mechanism for the emergence of a causal structure, or even a distinct dimension that could be associated with time.
Indeed, it is an often disputed claim that the discrete structure of the lattice prevents the model exhibiting Lorentz covariance as the $SO(3,1)$ symmetry is broken by the lattice.
We do not believe that this is a fatal flaw in discrete approaches and refer the reader to the work of Amelino-Camelia \cite{amelino2002relativity} and the comprehensive review by Hossenfelder \cite{hossenfelder2013minimal} for the details on how discrete geometries and Lorentz covariance can be reconciled.
Further, despite the simplicity and familiarity of the underlying physical model, we must consider the model and its refinements as essentially a ground up proposal for a model of emergent spacetime with no explicit connection to the well tested theories of QM and GR.
For these proposals to be taken seriously this connection with GR in particular is essential if any experimentally testable consequences are to be proposed.

It is to that question we will address ourselves in this paper.
The key insight will be to leverage recent results in the study of discrete curvature \cite{tee2021enhanced} that will allow us to analyze directly the Hamiltonian formulation of GR in a discrete setting.
We will seek to illustrate that by reinterpreting the emerged geometry as spatial, the Hamiltonians for the IG models can be inferred from a discretized formulation of Hamiltonian or Canonical GR (CGR), at least up to a choice of gauge or spacetime foliation.
In essence we decompose the spacetime manifold into $M=G(V,E) \bigotimes \mathbb{R}$, with $G(V,E)$ being the emerged spatial graph, and time $t \in \mathbb{R}$ being regarded as the sequence of labels on the state of the spatial graph.
It should be noted that there is no requirement on time being continuous, and we could equally well have chosen $M=G(V,E) \bigotimes \mathbb{Z}$.
Our argument rests upon the revolution in recent years in discrete  differential geometry that now provides a rich framework of curvature measures analogous to the familiar Riemann-Ricci  curvature of a smooth manifold.
This allows us to identify the induced curvature of a spatial slice with the discrete curvature of a spacetime graph, and the interpretation of spin-spin interactions as a kinetic energy term.
There are two options for discrete curvature that we will consider,the Forman-Ricci (FR) \cite{forman2003bochner} and Ollivier-Ricci (OR) \cite{ollivier2011visual} curvature.
It has been suggested that OR curvature has as its continuum limit \cite{van2021ollivier} the normal Ricci curvature of a manifold if one considers the graph as embedded in it.
It has however the drawback of being difficult to work with and does not have the supporting structure of index theorems, differential forms and vector fields that FR curvature possesses.
However, in a  recent result \cite{tee2021enhanced} it is proven that for certain important classes of graphs FR and OR curvature are simply related, allowing us to work with FR curvature and assume that in low energy continuum limit of the emerged geometry, we can recover the normal Ricci curvature.
Our claim in this paper is that starting with the Hamiltonian formulation of GR, we can propose a model that is broadly equivalent to the IG models, implying in turn that it may be possible to claim that such models have GR as a low energy limit.
We refer to the model derived from CGR as the Canonical Ising Model (CIM).

To investigate the properties of the CIM model we use numerical simulations to directly compute the ground states using  Glauber dynamics \cite{muller2012neural}, and present the results.
We will show, in the case of the FR curvature, that the ground states are broadly comparable to those produced in the IG models, specifically the QMD version, having a stable and highly local geometry. 
The OR curvature model though is more problematic.
Drawing on the result in \cite{tee2021enhanced} to leading order the Hamiltonian for FR curvature is a polynomial in the average degree $k$, which for a given value of the coupling constant has a well defined minimum energy at a specific value of $k$.
In the case of OR curvature, however, the leading order approximation of the Hamiltonian is linear in $k$ and so does not have such a value.
The existence of a stable energy minimum requires the insertion of a power of $k$, which is precisely the equivalence of the two curvatures established in \cite{tee2021enhanced}.

Focusing on the ground states of the CIM using Forman curvature, we see that the ground states of the graphs mirror the QMD models at least down to $3$ dimensions, below which the graphs show very divergent behavior.
This includes an explosion in clustering, and divergence in both intrinsic and extrinsic measures of dimensionality, which indicates that there may be distinct phases in the topology of the ground states of the CIM model.
This phase transition could perhaps point to something special about the model at three spatial dimensions, corresponding to the familiar observed $3+1$ dimensionality of the Universe.
Specifically, when we compute both the intrinsic and extrinsic curvature of the ground state graphs using the techniques described in \cite{jonsson1992intrinsic,ambjorn1997quantum}, we see that there is a distinct divergence of dimension measures below a critical value that lies between $3$ and $4$.
The CIM ground state graphs have a chaotic and `crumpled' topology below this, although the precision of the dimension at which this occurs is not sharp due to the computational limits on the size of the graphs we can numerically simulate. 
We also note a persistent negative spatial curvature of the model, which tends towards zero as dimension reduces.
The constraints from the canonical treatment of GR indicate that this negative spatial curvature is balanced by the extrinsic curvature and  may provide a hint of the global topology of discrete spacetime.

\subsection{Organization of this paper}

We begin our treatment in the next section \Cref{ssec:qmd} with an overview of the models previously studied, to enable detailed later comparison to the new discrete CIM model.
In particular we focus upon the QMD model introduced in \cite{tee2020dynamics}, as it has a simple Hamiltonian including a correction term to suppress triangles.
This correction term will prove to be important and can be reinterpreted as approximating discrete curvature in the Hamiltonian.

We present an overview of canonical GR, for completeness in \Cref{sec:discreteGR}, sufficient to be able to propose a discrete canonical model.
We operate in a specific choice of gauge, the `Gaussian' gauge, which simplifies the Hamiltonian.
Using these results, we briefly survey discrete approaches to curvature in \Cref{sec:curvature}, focusing on the Forman-Ricci and Ollivier-Ricci treatments.
To use these in the our discrete model we need certain mean-field approximations that we outline in \Cref{ssec:mean-field}.
We bring all of the elements together to propose our discrete canonical Ising model (CIM), in \Cref{ssec:discrete_hamiltonian}, using the FR curvature, and discuss how this compares to Ising models such as QMD.
We are then able to show analytically that the QMD Hamiltonian is, to first order in certain mean-field parameters, an approximation to CIM.

To explore the properties of the CIM model we describe the numerical computation of the ground states of this model in \Cref{sec:results}, and compare these to QMD.
We include the connectivity and locality properties of the models, but also investigate some topological properties not previously studied.
These include the discrete curvature and density of chordless cycles in the graph up to length $5$.
Additionally, it has also been proposed \cite{trugenberger2015quantum,tee2020dynamics} that the free parameter in the model, the coupling constant $g$, `runs' with energy.
As such we can consider the states of the model at different values of $g$ representing the evolution of the geometry from an origin event.
We discuss how our results could provide an alternative view of a `topological big bang' with spatial expansion that is initially rapid, slowing down as the spatial dimension of the model approaches $3$.

Finally in \Cref{sec:conclusion} we conclude with an assessment of the merits and deficiencies of the model we have proposed, and highlight potential further lines of inquiry.

\subsection{Overview of Ising emergent geometry}
\label{ssec:qmd}
The model originally introduced by Trugenberger \cite{trugenberger2015quantum} was proposed as an Ising model, with spin `qubits' located on the vertices of a graph, and edges being favored between spins that align due to their ferromagnetic interaction.
Operating in opposition to edge formation is a link frustration term, expressed as an anti-ferromagnetic Ising interaction between spin states defined on the edges.

The statement of the model \cite{tee2020dynamics} can be made precise by defining the Hilbert spaces for the edges and the $N$ vertices of a simple, undirected graph $G(V,E)$ where $V$ is the set of vertices and $E \subset \{V \times V \}$ the edges connecting any two distinct vertices.
It should be noted that throughout this paper we make use of standard graph theory notation consistent with standard texts such as Bollobas \cite{bollobas2013modern}.
On each vertex $v_i $ we associate the Hilbert space $\mathcal{H}_{i} = \mbox{span} \{ \ket{i,0}, \ket{i,1} \}$, and define a spin operator obeying $\hat{s}_{i} \ket{i,s}=s_i\ket{i,s}$ on this space.
We assert the usual anti-commutation relations and ladder operators consistent with spin $1/2$ fermions,
\begin{align}
	\{ \hat{s}^{+}_i, \hat{s}^{+}_j \},  \{ \hat{s}^{-}_i, \hat{s}^{-}_j \} = 0, \{ \hat{s}^{-}_i, \hat{s}^{+}_j \}&= \delta_{ij} \mbox{,} \\
	\hat{s}^{+}_i \ket{ i,1} = 0, \hat{s}^{+}_i \ket{ i,0} &= \ket{i,1} \mbox{,} \\
	\hat{s}^{-}_i \ket{ i,1} = \ket{i,0}, \hat{s}^{-}_i \ket{ i,0} &= 0 \mbox{.}
\end{align}

For the edges we proceed similarly with $\frac{1}{2}N(N-1)$ Hilbert spaces,  $\mathcal{H}_{ij} = \mbox{span} \{ \ket{i,j,0}, \ket{i,j,1} \}$, with the state $\ket{i,j,1}$ indicating the presence of an edge, and $\ket{i,j,0}$ its absence.
In a parallel fashion to the vertices we define fermionic edge annihilation and creation operators,

\begin{align}
	\{ \hat{a}^{\dagger}_{ij}, \hat{a}^{\dagger}_{kl} \},  \{ \hat{a}_{ij}, \hat{a}_{kl} \} = 0, \{ \hat{a}_{ij}, \hat{a}^{\dagger}_{kl} \}&= \delta_{ik} \delta_{jl} \mbox{,}\\
	\hat{a}^{\dagger}_{ij} \ket{ i, j, 1} = 0, \hat{a}^{\dagger}_{ij} \ket{ i, j, 0} &= \ket{i, j, 1} \mbox{,}\\
	\hat{a}_{ij} \ket{ i, j, 1} = \ket{i,j, 0}, \hat{a}_{ij} \ket{ i, j, 0} &= 0 \mbox{.}
\end{align}

The adjacency matrix of the graph $A_{ij}$ (defined as having the value $1$ when there is an edge between vertices $i$ and $j$, and zeros elsewhere and $A_{ii}=0$), can be represented using these edge annihilation/creation operators as $A_{ij} = \hat{a}^{\dagger}_{ij} \hat{a}_{ij}$.
With these definitions we are able to state the Hamiltonian of the model, termed the  Dynamical Graph Model (DGM) described in  \cite{trugenberger2015quantum} as,

\begin{equation}\label{eqn:trugenberger}
	H_{DGM}=\frac{g^2}{2} \Bigg ( \sum\limits_{i \neq j}^N \sum\limits_{ k \neq i,j }^N A_{ik}A_{kj}  \Bigg ) - \frac{g}{2}  \sum\limits_{i,j} s_i A_{ij}  s_j \mbox{.}
\end{equation}

It is possible using numerical methods solve for the ground state graphs of this Hamiltonian at different values of $g$, and it was found that they possess certain attractive features:

\begin{description}
    \item [Regular, Euclidean flat ground state] The ground state corresponds to a regular graph where nearly all of the nodes posses the same average degree (number of incident edges upon a given vertex) $k$. This configuration is referred to as a `large world' in network science, which means that the graph has a high degree of locality with few `short-cuts' between distant nodes.
    \item [Low dimensionality] There are several ways to quantify the dimension of a graph, both intrinsic dimension (as would be measured by an observer confined to the graph), and extrinsic dimension measured from the point of view of a higher dimensional space in which the graph is embedded \cite{ambjorn1997quantum}. These two measures agree for the ground state until around $d=4$, at which point the extrinsic dimension does not reduce any further. This points to a `preferred' low dimension for the graph.
    \item [Entropy area law] It is possible to demonstrate that the ground state graph possesses a measure of informational entropy which is related to the size of the boundary of the graph or any defects in it, rather than its bulk.
\end{description}

Despite these attractive features, the ground state obtained using Eq. \eqref{eqn:trugenberger} has persistent non-zero clustering. 
The presence of clustering is highly undesirable as it amounts to a loss of locality in the emerged geometry.
Accordingly, the model was further refined by the insertion of perturbation terms in the third and fourth powers of the adjacency matrix which penalize triangles and favors squares by Trugenberger \cite{trugenberger2016random}, but it was later demonstrated that an even simpler Hamiltonian with one dimensionless coupling constant could achieve the same result without the use of additional parameters \cite{tee2020dynamics}.
This model is referred to as Quantum Mesh Dynamics (QMD), as it formed the basis of a dynamical theory of matter in the emerged ground state.
The Hamiltonian proposed in this model is,

\begin{equation}\label{eqn:qmd_hamiltonian}
	H_{QMD}=\frac{g^2}{2} \Bigg ( \Tr (A^3) +  \sum\limits_{i \neq j}^N \sum\limits_{ k \neq i,j }^N A_{ik}A_{kj}  \Bigg ) - \frac{g}{2}  \sum\limits_{i,j} s_i A_{ij}  s_j \mbox{.}
\end{equation}

An intriguing feature of this Hamiltonian, and those proposed as extensions to DGM \cite{trugenberger2016random}, is the appearance of sequential powers of the adjacency matrix, which could suggest that they are an approximation to a more fundamental model.
In particular, do these powers of the adjacency matrix suggest that the Hamiltonian is in fact comprised of a kinetic term relating to neighbor spin interactions and a potential term relating to the structure and topology of the spacetime graph?
We will show in the following sections that indeed this could be the case, and the first term is an approximation of the spatial curvature of the graph.
When decomposed in this fashion we have a Hamiltonian that is strongly analogous to canonical GR, at least up to the choice of `lapse' and `shift' vectors.
The identification of $\sum\limits_{i,j} s_i A_{ij}  s_j$ as the kinetic term was investigated in \cite{tee2020dynamics,tee2021quantum}, when one considers matter as excited defects in the ground state of the emerged geometry. 
In that specific instance the precise formulation uses spin ladder operators and the graph Laplacian, but in form at least we consider the second term to be kinetic.
It is these observations, along with the numerical evidence presented, that underpins the central contribution of this work that IG models and canonical GR are intimately related.
This result is intriguing as it could point to the possibility of GR emerging as a continuum limit to an emergent Ising geometry theory, similar to the models described in this paper.

\section{Discretizing Canonical Gravity}
\label{sec:discreteGR}
Our starting point is the the Hamiltonian formulation of GR \cite{arnowitt1959dynamical,arnowitt2008republication,poisson2004relativist}, used as the basis for the canonical approach to quantum gravity \cite{dewitt1967quantum1,dewitt1967quantum2,dewitt1967quantum3}.
This approach to GR is well documented in the standard texts cited, and here we provide the minimum overview necessary to frame our proposed Hamiltonian.

Hamiltonian GR relies upon a decomposition of spacetime into a series of spacelike foliations $\Sigma_t$, at constant time $t$.
For the following discussion Greek indices $\alpha,\beta$ run from $0$ to $4$ and Latin indices $i,j,k$ from $1$ to $3$, and we will work in the `East Coast' $(-1,1,1,1)$ metric signature.
The foliation induces a metric on the hypersurfaces, which we denote as $h_{ij}$ to distinguish it from the $4$-metric $g_{\mu \nu}$.
We can decompose the $4$-metric into a combination of the induced metric $h_{ij}$ and two new quantities, the lapse $N$, and shift vector $N^i$, which we define below.

On the hypersurfaces we provide a local coordinate system $y^i$, and a transformation $x^{\alpha}=x^{\alpha}(t,y^i)$ that relates the local coordinates at  time $t$ of the hypersurface to the general coordinates of spacetime.
The physical interpretation of lapse and shift vectors can be understood in terms of a congruence of curves $\gamma$ that intersect each spacelike foliation, such that the local coordinates $y^i$ of each intersection with each hypersurface are preserved.
The transformation from local to global coordinates at constant $t$, $e^{\alpha}_i=\left ( \pdv{x^{\alpha}}{y^i} \right )_t$, defines a natural set of basis vectors in the hypersurface.
Using these basis vectors, if $n^{\alpha}$ is the normal vector to the hypersurface, the tangent vector to a curve $\gamma$, $v^{\alpha}$ can be written as $v^{\alpha} = N n^{\alpha} + N^{a} e^{\alpha}_a$.
It is important to underline that the choice of $N$ and $N^a$ is arbitrary and can be thought of as a choice of gauge for a given foliation.
This freedom in the choice of lapse and shift vector means that they can not be dynamic variables of the theory, and this results in certain constraints on the solutions of the resultant equations of motion often referred to as the Hamiltonian and momentum constraints. 

In terms of the induced metric, lapse and shift vector the $4$-metric can be decomposed as,
\begin{align}
   g_{\mu \nu} &= \begin{pmatrix} 
                -N^2 + N_i N^i & N_i \\
                N_j & h_{ij}
                \end{pmatrix} \text{,} \label{eqn:decomposed_metric_cv}\\ 
    g^{\mu \nu} &= \begin{pmatrix}
                    -N^{-2} & N^{-2} N^{i} \\
                    N^{-2} N^j & h^{ij} - N^{-2} N^i N^j
                  \end{pmatrix} \label{eqn:decomposed_metric_cont}\mbox{.}
\end{align}
Along with these definitions, it is then possible to define both an intrinsic and an extrinsic  curvature that can be related back to the full Riemann tensor for spacetime using the Gauss-Codazzi equations.
The intrinsic tensor $~^3\!R_{ij}$ is interpreted as the curvature as experienced in the hypersurface, and  the extrinsic curvature $K_{ij}$,  the curvature of the foliations as embedded in the full spacetime.

In what follows we draw largely upon the standard text by Poisson \cite{poisson2004relativist}.
It is possible to express the theory of GR in terms of a Hamiltonian where the dynamic variables are the induced metric $h_{ij}$ and its time derivative $\dot{h}_{ij}  = \partial_t h_{ij}$.
The canonical momentum $p^{ij}$ can be defined using the gravitational Lagrangian density $\mathcal{L}_G$ as,

\begin{equation}
    p^{ij} = \pdv{ \left( \sqrt{ -g} \mathcal{L}_g \right ) }{\dot{h}_{ij}} = \frac{\sqrt{h}}{16 \pi } (K^{ij} -K h^{ij}) \mbox{,}
\end{equation}
where $K=h_{ij}K^{ij}$ is the extrinsic curvature scalar.

Using this expression for the canonical momentum we state the following expression for the gravitational Hamiltonian $H_G$,
\begin{equation}\label{eqn:gr_ham}
\begin{split}
    (16 \pi ) H_G &= \int_{\Sigma_t} \left [ N( K^{ij} K_{ij} -K^2 - ~^3 R) -2 N_i (K^{ij} - K h^{ij})_{|j} \right ] \sqrt{h} ~ d^3 y\\
    &\mbox{ + surface terms, }
\end{split}
\end{equation}
where $(\dots)_{|j}$ refers to induced metric, $h_{ij}$ compatible covariant differentiation.
We can express \Cref{eqn:gr_ham} in terms of the momentum $p^{ij}$ and its scalar contraction $p=h_{ij}p^{ij}$ as follows,

\begin{equation}
\begin{split}
    (16 \pi ) H_G &= \int_{\Sigma_t} \left [ N \left ( \frac{(16 \pi )^2}{h} ( p^{ij}p_{ij} -\frac{1}{2}p^2 ) - ~^3 R \right ) -2N_i \left ( \frac{16 \pi }{\sqrt{h}}p^{ij} \right )_{|j}  \right ] \sqrt{h} ~d^3 y\\
    &\mbox{ + surface terms. }
\end{split}
\end{equation}
We now invoke the freedom to choose the value of $N$ and $N_a$ by setting $N=1$, and $N_a=0$, sometimes referred to as the ``Gaussian gauge''.
On each foliation $\Sigma_t$ with this choice of lapse and shift vector, one obtains a system of Gaussian normal coordinates, for which observers  with $x^i = $ constant represent local surfaces of simultaneity (see \cite{thorne2000gravitation} sections $21,27$).
Alternatively, one can also think of each curve $\gamma$ intersecting each hypersurface orthogonally.
Discarding surface terms we are left with,
\begin{equation}\label{eqn:H_can}
     (16 \pi ) H_G = \int_{\Sigma_t} \left [ \frac{(16 \pi )^2}{h} ( p^{ij}p_{ij} -\frac{1}{2}p^2 ) -~^3R \right ] \sqrt{h} ~d^3 y
\end{equation}
In the standard interpretation of classical mechanics $H=T+V$, where $T$ is the kinetic energy and $V$ the potential, and indeed we see in this Hamiltonian that there is a term in the square of the canonical momenta that we can equate to $T$.
Making this association we interpret $(p^{ab}p_{ab}-p^2)$ as our kinetic energy $T$ and $(-~^3R)$ as the potential energy $V$, up to proportionality constants.
Our strategy is to replace these terms with their discrete analogs and propose a Hamiltonian that represents a discretized form of the CGR Hamiltonian $H_G$. 
For the intrinsic curvature it is natural to substitute the discrete curvature of a graph that we will introduce in \Cref{sec:curvature}, and denote as $\kappa_{ij}$.
Discrete curvature in the context of a graph is defined on the edges, and the equivalent of the scalar curvature $~^3R$ defined at a point, is the sum over all edges incident at a given vertex $\sum\limits_{j \sim i} \kappa_{ij}$, with the notation $j \sim i$ indicating that the vertex $j$ is connected by an edge to the vertex $i$.
In this way we identify vertices of the graph with points in spacetime in the continuum limit.
We will make this much more precise in \Cref{sec:curvature}. 
So much for the intrinsic curvature and $V$, but what can play the role of kinetic energy term?

In \cite{tee2020dynamics} it was demonstrated that a particular choice of interaction Hamiltonian between opposite spins in the mesh was consistent with the non-relativistic wave equation in the continuum limit. 
This required an interaction Hamiltonian of the form,

\begin{equation}\label{eqn:H_mod}
	\hat{H}_{ij}^{I} = -\frac{g}{2\epsilon_m r^2_{ij} }\hat{s}_i^{+} ( 1 + L_{ij} ) \hat{s}_j^{-} \text{,}
\end{equation}
where $L_{ij}$ is the Laplacian matrix of the graph, defined as $L_{ij} = \Delta_{ij} -A_{ij}$, where $\Delta_{ij} = \delta^i_j k_i$ is the degree matrix.
For the remaining terms, $r_{ij}$ is the distance (i.e. shortest path length) between vertices $i$ and $j$, $\epsilon_m$ the excitation energy necessary to flip a spin, $\hat{s}^{\pm}_{i}$ the spin ladder operators at a given vertex, and $g$ the coupling constant.
It is precisely $L_{ij}$ that in the continuum limit gives rise to the momentum term, arising from the fact that $L_{ij}$ is associated with $-\nabla^2$ when discrete graph models are taken to a continuum limit \cite{chung1997spectral}.
The Laplacian compares how much of a particular edge, incident at a vertex, represents of the total connectivity of that vertex.
In an unweighted graph this is of course evenly distributed amongst all edges.
It is this property that in discrete dynamical models measure the `flow' of relative influence in a particular direction in a graph away from a specific vertex.
The structure of this interaction Hamiltonian models the dynamics of defect excitations in the ground state of IG models.
Fundamentally it compares spins at different vertices in the graph, and in the ground state this structure is captured by the adjacency matrix.
In fact, terms involving simply the adjacency matrix and vertex spin interactions such as $\sum\limits_{ij} \hat{s}_i A_{ij} \hat{s}_j$ are often associated as representing the `kinetic' energy of the Ising model when the model evolves in time \cite{creutz1986deterministic,marcjasz2017phase}.
In fact the Hamiltonians of the DGM and QMD models, \Cref{eqn:qmd_hamiltonian} and \Cref{eqn:trugenberger}, both contain such a term, capturing the tendency for aligned spins to generate an edge and therefore `propagate' influence in the graph.

Pursuing this analogy, we identify this $\sum\limits_{ij} \hat{s}_i A_{ij} \hat{s}_j$ term as representing `momentum' for the purposes of constructing our discrete Hamiltonian as realized in the graph $G(V,E)$.
To propose this Hamiltonian we therefore make the following substitutions in Eq. \eqref{eqn:H_can}, 
\begin{align}
    \int_{\Sigma_t} \dots \sqrt{h} ~ d^3 y & \rightarrow l_p^3 \sum\limits_{i \in V} \mbox{;} \\ 
    \frac{(16 \pi )^2}{h}(p^{ab}p_{ab} -\frac{1}{2}p^2) & \rightarrow -\frac{g}{2} \sum\limits_{i,j \in V} \hat{s}_i A_{ij} \hat{s}_j \mbox{;}  \label{eqn:laplace_sub}\\
    ~^3R(y) & \rightarrow  \sum\limits_{j \sim i} \kappa_{ij}  \mbox{.} 
\end{align}
In the first substitution we have added in the $l^3_p$ factor, with the assumption that $l_p$ is the lattice spacing, for completeness.
This value arises from the discretization of the measure and as it has no effect on our numerical analysis, and will  be disregarded in what follows.

We have introduced a coupling constant $-\frac{g}{2}$ into our model, and although we could absorb any proportionality for the potential term into this one coupling constant term in the Hamiltonian, we will use $-\frac{g}{2}$ for the kinetic term and its square $\frac{g^2}{4}$ for the potential term to allow us to balance between the two contributions when we simulate.
To be explicit, the simulations use $g$ as the free parameter of the model, which ultimately we will see is linked to the dimensionality of spacetime.
In our simulations we vary the value of $g$, and we find that for our Hamiltonian the range $0.04 < g < 0.14$ yields ground states that are particularly interesting.
As $g$ increases the ground state is obtained when the curvature potential energy balances the spin-spin kinetic energy at the minimum of the Hamiltonian.
For large values of $g$ the curvature term required to balance the kinetic contribution and produce this minimum can be smaller, which we shall see corresponds to a ground state graph that has an overall lower connectivity and more closely resembles a square lattice.
We can now finalize the prescription for obtaining a discretized CGR model, and propose our canonical Ising Hamiltonian $H_{CIM}$ as,

\begin{equation}\label{eqn:h_comb}
    H_{CIM} = -\frac{g}{2}\sum\limits_i \sum\limits_j \hat{s}_i A_{ij} \hat{s}_j - \frac{g^2}{4} \sum\limits_i \sum\limits_j \kappa_{ij}\text{.}
\end{equation}
It will be noted that the negative sign in the $-^3R$ term carries over into the above expression to give an overall negative sign to the term in discrete curvature.

In order to make use of Eq. \eqref{eqn:h_comb}, we need an expression for the discrete curvature of the spacetime graph.
We turn to the computation of this in \Cref{sec:curvature}.

\section{Discrete Curvature and the Discrete Canonical Hamiltonian}
\label{sec:curvature}
We will present a very brief overview in this section of two popular methods of computing graph curvature, the Forman-Ricci  and Ollivier-Ricci  curvature.
The principle aim of this section is to obtain an expression for the curvature in terms of graph primitives such as the adjacency matrix.
We will draw upon recent results regarding the mean field values of these curvatures \cite{tee2021enhanced}, and do not intend this to be a comprehensive treatment of what is a complex subject.

\subsection{Forman-Ricci}
\label{ssec:forman}
We begin with a brief overview of FR curvature, originally proposed by Robin Forman, as a measure of curvature on CW (closure-finite, weak topology) complexes  \cite{forman2002combinatorial,forman2003bochner,forman2004topics}.
The work is of particular interest for applications to physics, as it includes the discrete analogs of differential forms, Morse theory and Ricci curvature to their counterparts in the differential geometry of smooth manifolds.
The definition of FR curvature draws upon an analogy to the identities developed  by Bochner \cite{bochner1946vector} regarding the decomposition of the Riemannian-Laplace operator defined on $\Omega^p(M)$, the space of $p$ forms of a manifold $M$.
In this decomposition, the Laplace operator can be expressed as a sum of the square of the covariant derivative plus a curvature term.
This is the origin of the Bochner-Weitzenb{\"o}ck identity and in its discrete form it is used to derive the formula for the FR curvature.

Forman's analysis is conducted using the formalism of CW complexes (the reader is referred to standard texts on algebraic topology for more detail such as Hatcher \cite{hatcher2002algebraic}).
The concept is  similar to simplicial complexes, which are perhaps more familiar due to the calculating theorems used to compute Homology and Homotopy groups of topological spaces \cite{nash1988topology,nakahara2003geometry}, but the formalism is more general.
The building blocks are the $p$-cells, where the $p$  refers to the dimension of the cell, and intuitively a $d$ dimensional CW complex built by gluing $p \leq d$ complexes along shared faces.
For example, a $0$-cell is a point $p_1$, a $1$-cell an edge between two points $\langle p_1 p_2\rangle$, and a $2$-cell could be a triangle, with the interior included,  bounded by three $1$-cells $\langle p_0 p_1 \rangle , \langle p_1 p_2 \rangle , 
\langle p_2 p_3 \rangle$, or indeed longer cycles including their interiors.
For our purposes we will focus on cell complexes up to $p=2$, which are essentially equivalent to graphs, with the addition that chordless cycles in the graph are assumed to bound a $2$-cell.
We will define more precisely the concept of chordless cycles in \Cref{ssec:mean-field}.

An important concept is the boundary of a $p$-cell, being the set of $p-1$ cells that contain the cell.
Intuitively for a $1$-cell $\langle p_0p_1 \rangle$, the boundary is the collection of points $p_0$ and $p_1$, and for a general $p-$cell, $\alpha_p$, we say it is a proper face of a $p+1$ cell $\beta$ if it is a member of the boundary set of $\beta$, and we write $\alpha_p < \beta_{p+1}$, or conversely $\beta_{p+1} > \alpha_p$.
A $p$-cell CW complex $M$, over $\mathbb{R}^p$, is then a collection of cells $\alpha_p p \in \{0,\dots,p\}$, with the restriction that any two cells are joined along a common proper face, and all faces of all complexes are contained in the cell.

The key concept used to defined the  curvature of cell complexes is the definition of the neighbors of complexes of arbitrary degree.
We reproduce the definition first stated by Forman here \cite{forman2002combinatorial,forman2003bochner},

\begin{definition}\label{def:neighbor}
	$\alpha_1$ and $\alpha_2$ are $p$-cells of a complex $M$. $\alpha_1$,$\alpha_2$ are neighbors if,
	\begin{enumerate}
		\item $\alpha_1$ and $\alpha_2$ share a $(p+1)$ cell $\beta$ such that $\beta > \alpha_1$ and $\beta > \alpha_2$, or
		\item $\alpha_1$ and $\alpha_2$ share a $(p-1)$ cell $\gamma$ such that $\gamma < \alpha_1$ and $\gamma < \alpha_2$.
	\end{enumerate}
\end{definition}

In addition, we can partition the set of neighbors of a complex into parallel and non-parallel neighbors.
We say that two $p$-cells $\alpha_1$,$\alpha_2$ are parallel, if one but not both of the conditions in \Cref{def:neighbor} are true, and we write $\alpha_1 \parallel \alpha_2$.

FR curvature is defined as a series of maps $\mathcal{F}_p : \alpha_p \rightarrow \mathbb{R}$, defined for each value of $p$.
In the case of an unweighted CW complex, this reduces to a simple form,
\begin{equation}\label{eqn:fr_simple}
	\mathcal{F}_p (\alpha_p) = \# \{ \beta_{(p+1)} > \alpha_p \} + \# \{ \gamma_{(p-1)} < \alpha_p \} - \# \{ \epsilon_p \parallel \alpha_p \} \mbox{,}.
\end{equation}
where $\epsilon_p$ is a $p$-cell that is a parallel neighbor of $\alpha_p$, and $\epsilon_q \neq \alpha_p$.
We can state this in words as the number of $(p-1)$-cells that bound $\alpha_p$, plus the number of $(p+1)$-cells that $\alpha_p$ is part of the boundary of, minus the number of $p$-cell parallel neighbors of $\alpha_p$.

The curvature function for $p=1$ is specifically identified as the analog of Ricci curvature, and defines the curvature of the edges of a graph, and we refer to this as the Forman-Ricci curvature.
We distinguish this particular value  of the FR curvature for $p=1$ by the notation $\kappa_{ij}^f = \mathcal{F}_1(e_{ij})$.
Fortunately for a regular mesh \Cref{eqn:fr_simple} is highly simplified, as the vertices and edges constitute the $0$ and $1$-cells, and we assume that any chordless cycles in the graph constitute the $2$-cells.
The interpretation of \Cref{eqn:fr_simple} is often  referred to as the `augmented' FR curvature.
It should be noted that many authors do not consider arbitrary length cycles, and only consider triangles.
This is not appropriate for our purposes as a flat square mesh would not have zero curvature.
Henceforth we will assume any reference to FR curvature refers to this augmented form.

At each vertex $v_i$, if $j \sim i$ indicates that the vertex $v_j$ is connected to $v_i$ by one edge, we define the analogy of the Ricci scalar curvature at $v_i$ as,

\begin{equation}\label{eqn:scalar_forman}
    \kappa^f(v_i) = \sum\limits_{j \sim i} \kappa^f_{ij} \text{.}
\end{equation}
For the whole graph, $G(V,E)$, we can sum over all vertices, to compute the total curvature as follows,
\begin{equation}\label{eqn:scalar_forman_graph}
    \kappa^f = \sum\limits_{i} \kappa^f(v_i) = 2\sum\limits_{i,j \in V} \kappa^f_{ij}  \mbox{,}
\end{equation}
with the factor of $2$ coming from the fact that the graph is undirected and each edge is counted twice.

\subsection{Ollivier-Ricci}
\label{ssec:ollivier}
The second form of curvature we consider is Ollivier-Ricci, first introduced by Yann Ollivier \cite{ollivier2007ricci,ollivier2009ricci,ollivier2011visual}.
This curvature has a close association to the conventional notion of curvature for a smooth manifold.
It is based upon the well known result from Riemannian geometry concerning small balls surrounding two points in a curved space.
Between the two points we consider a geodesic, and in flat space the average geodesic distance between all other points in the two balls is the same as the geodesic distance between the centers.
Positive curvature shortens this average distance and negative curvature lengthens it.

The starting point for OR curvature is to define the graph to be a metric space, with the metric provided by graph distance, and the role of the balls is replaced by unit normalized probability distributions on neighboring vertices.
Wasserstein distance between `unit balls' centered around two points $i$ and $j$, denoted $b_i, b_j$ is used to define geodesic distance between $i$ and $j$.
This relies upon a unit normalized measure $\mu_i$ of `mass', and a transference plan that exchanges this `mass' between distributions on $b_i$ and $b_j$.
Wasserstein distance is then defined to be the optimal such transport plan.

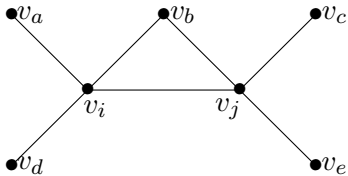
\begin{figure}[htbp]
    \centering
    \begin{tikzpicture}[node distance=0.5cm]
	    \node [black] at (0,2) {\textbullet};
	    \node [black] at (1,1) {\textbullet};
	    \node [black] at (0,0) {\textbullet};
	    \node [black] at (2,2) {\textbullet};
	    \node [black] at (3,1) {\textbullet};
	    \node [black] at (4,2) {\textbullet};
	    \node [black] at (4,0) {\textbullet};
	    
	    \draw [-] (0,0) -- (1,1);
	    \draw [-] (0,2) -- (1,1);
	    \draw [-] (1,1) -- (3,1);
	    \draw [-] (1,1) -- (2,2);
	    \draw [-] (2,2) -- (3,1);
	    \draw [-] (3,1) -- (4,2);
	    \draw [-] (3,1) -- (4,0);

	    \node at (1.1,0.75) {$v_i$};
	    \node at (2.85,0.75) {$v_j$};
	    \node at (0.25,2) {$v_a$};
	    \node at (2.25,2) {$v_b$};
	    \node at (4.25,2) {$v_c$};
	    \node at (0.25,0) {$v_d$};
	    \node at (4.25,0) {$v_e$};
\end{tikzpicture}
\caption{An example graph, upon which we will compute the OR curvature of the edge $e_{ij}$. For this graph $b_i=\{v_i, v_j, v_a,v_b,v_d\}$ and $b_j=\{v_j, v_i,v_b,v_c,v_e\}$.}
\label{fig:or_example}
\end{figure}

For a graph $G(V,E)$, with the graph distance metric, these concepts can be made more precise as follows.
The unit ball $b_i$ of a node $v_i$ corresponds to the set of nodes, including $v_i$ that are connected by one edge to $v_i$, and the graph distance metric $d(v_i,v_j)$ is the shortest path between two vertices $v_i$ and $v_j$.
Our metric space is therefore defined as $M=( G, d( v_i,v_j) )$.
The probability measure we use is called the graph counting measure, defined using the fractional cardinality of any subset of vertices $X \subset V$, simply stated as $\mu(X) = |X|/|V|$.
Applied to our unit balls $b_i$ we have $\mu(v_x^i)= 1 / |b_i|$, for $v_x^i \subset b_i$.
This obeys the sum to unity constraint of a probability measure $\sum_a \mu( v_x^i ) = 1$, and we note for a node of degree $k_i$ that $\mu( v_x^i ) = 1/(k_i+1)$ is trivially valid as such a measure.
To obtain an optimal transport plan we need to calculate the minimum cost to carry the probability distribution from $b_i$ to $b_j$. 
This is not a simple undertaking, but for regular graphs it is simplified by the symmetry of the neighborhoods of any given vertex.
As an example, in \Cref{fig:or_example} we depict such a graph, and the unit balls $b_i,b_j$ that form the neighborhood of $v_i$ and $v_j$ at a graph distance of $1$.
The transport cost $\pi(v_x,v_y)$ for each vertex $v_x^i$ in $b_i$ is the distance the node has to `transport' its probability $\mu(v_x^i)$, or a portion of it, to a node $v_y^j$ in $b_j$, multiplied by the value of the portion of the probability distribution transported.
The objective of the transport being to swap the probability distributions surrounding the source and destination nodes, which in turn may involve transfers of probability distributions between all possible combinations of nodes.
For such a transport we have a cost of $\pi(v_x,v_y) = \mu(v_x^i) d( v_x, v_y )$.
For example, the node $v_a$ has a  value of $\mu=1/5$, and it needs to move a distance of $3$ hops to $v_c$, yielding a cost of $3/5$.
We need to do the same with $v_d$, but the nodes $v_b$, $v_i$ and $v_j$ in $b_i$ do not need to move.
The specific collection of such moves is a transport plan, and the transference $\xi(b_i,b_j)$ between $v_i$ and $v_j$ is then the sum of these transport costs for a given transport plan $\Pi(b_i,b_j)$, formally,
\begin{equation}
    \xi(b_i,b_j) =  \sum\limits_{x \in b_i, y \in b_j, \pi( v_x, v_y ) \in \Pi(b_i,b_j)} \pi( v_x, v_y ) \text{.}
\end{equation}
In the case of our example transference plan it is $6/5$, but in general there are a combinatorially large set of possible transports and therefore transport plans.
We are now ready to define the Wasserstein distance as the infinum of the transference over all possible transport plans,  
\begin{equation}
    W(b_i,b_j)= \inf\limits_{\Pi(b_i,b_j)} \xi( b_i,b_j) \text{.}
\end{equation}
Using this definition, the OR curvature $\kappa^o_{ij}$ of a given edge between $v_i$ and $v_j$ is then defined as,

\begin{equation}
    \kappa^o_{ij} = 1 - \frac{ W(b_i,b_j) }{ d(v_i,v_j )} \text{.}
\end{equation}

In the case of our graph in Fig . \ref{fig:or_example} the OR curvature of the edge $e_{ij}$ connecting $v_i$ and $v_j$ is therefore $\kappa^o_{ij} = (1-6/5) = -0.2$.

For an arbitrary graph without a high degree of symmetry and regularity, this computation is extremely involved and computationally expensive.
There are some exact calculations in closed form available \cite{kelly2019self}, but they rely upon some simplifying assumptions which may not hold in general.
However, for a regular graph, such as those known to be the ground states of the Ising emergent geometry models, these assumptions can be shown to hold.
We refer the reader to the cited literature for details.

\subsection{Mean-field results for discrete curvature}
\label{ssec:mean-field}
\begin{figure}[htbp]
    \centering
    \begin{tikzpicture}[node distance=0.5cm]
	    \node [black] at (0,0) {\textbullet};
	    \node [black] at (0,1) {\textbullet};
	    \node [black] at (1,0) {\textbullet};
	    \node [black] at (1,1) {\textbullet};
	    
	    \node [black] at (4,0) {\textbullet};
	    \node [black] at (4,1) {\textbullet};
	    \node [black] at (5,0) {\textbullet};
	    \node [black] at (5,1) {\textbullet};
	    
	    \draw [-] (0,0) -- (0,1);
	    \draw [-] (0,1) -- (1,1);
	    \draw [-] (1,1) -- (1,0);
	    \draw [-] (1,0) -- (0,0);
	    
	    \draw [-] (4,0) -- (4,1);
	    \draw [-] (4,1) -- (5,1);
	    \draw [-] (5,1) -- (5,0);
	    \draw [-] (5,0) -- (4,0);
	    \draw [-,red] (4,0) -- (5,1);

	    \node at (0,-0.25) {$v_1$};
	    \node at (1,-0.25) {$v_4$};
	    \node at (0,1.25)  {$v_2$};
	    \node at (1,1.25)  {$v_3$};
	    \node[text width=3cm] at (1.0,-1.0){Chordless cycle on $(v_1,v_2,v_3,v_4)$};
	    
	    \node at (4,-0.25) {$v_1$};
	    \node at (5,-0.25) {$v_4$};
	    \node at (4,1.25)  {$v_2$};
	    \node at (5,1.25)  {$v_3$};
	    \node[text width=3cm] at (5.0,-1.0){$(v_1,v_3)$ Bisects cycle creating two Chordless cycles $(v_1,v_2,v_3),~(v_1,v_3,v_4)$};
\end{tikzpicture}
\caption{We depict here two cycles of length $4$ involving the vertices $(v_1,v_2,v_3,v_4)$. On the left hand side the cycle is chordless and would represent a proper $2$-cell in a CW complex for the computation of FR curvature or a square in the computation of OR curvature. The cycle on the right hand side however is bisected by the edge between $v_1$ and $v_3$. As such it would contribute two triangles to any curvature computation, and zero squares.}
\label{fig:chordless}
\end{figure}
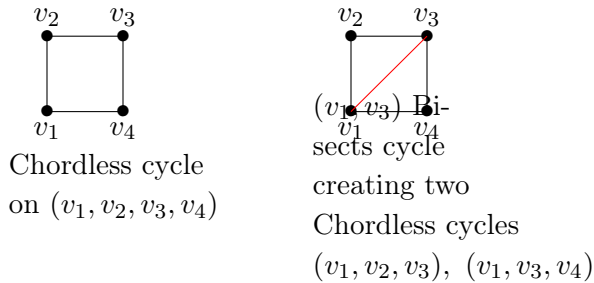

Fortunately the expressions for discrete curvature simplify dramatically when we constrain the types of graphs that we consider.
In particular the frequency of chordless cycles is critical.
A chordless cycles is used to represent the $2$-cells in the FR treatment of the graph, and indeed to compute the approximate values of OR curvature as described in \cite{kelly2019self,tee2021enhanced}.
Firstly let us define a chordless cycle as a closed path in the graph that contains no bisections.
More precisely, for a cycle of length $n$, being a collection of $n$ vertices connected path-wise by $n$ edges, it is a chordless cycle if and only if each vertex has only two neighbors considering only the vertices included in the cycle.
We present in \Cref{fig:chordless} two cycles of length $n=4$, to illustrate the difference.

\begin{figure}[htbp]
	\centering
	\includegraphics[scale=0.43]{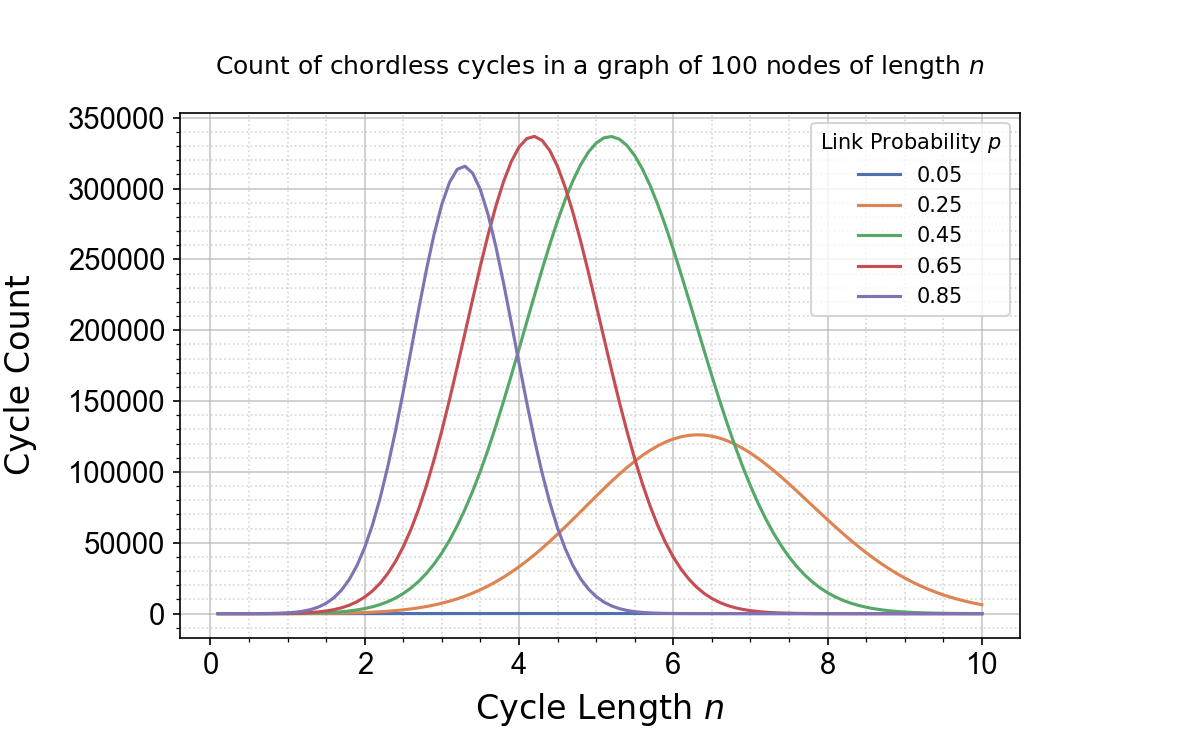}
	\caption{For $N=100$ vertices we generate random Gilbert graphs \cite{barabasi2016network}, varying the link probability $p$. For each value of $p$ we compute the number of chordless cycles of length $n$, plotting the theoretical value for this number.}
	\label{fig:chordless_cycles}
\end{figure}

Chordless cycles in a graph become increasingly rare as the length of the cycle increases.
This is easy to demonstrate for random graphs by noting that the link probability $p$, also defines the probability of a link not existing as $(1-p)$. 
Given that in a cycle of length $n$ there are $n$ links present out of a possible $\frac{1}{2}n(n-1)$ available, the cycle is therefore chordless with probability $(1-p)^{\frac{1}{2}n(n-3)}$.
We denote the number of chordless cycles of length $n$ by $\square_n$, and cycles of length $n$ that are either chordless or not by $\boxtimes_n$.
For a graph of $N$ vertices we have,

\begin{align}
    \square_n&=(1-p)^{\frac{1}{2}n(n-3)} \boxtimes_n \text{,and} \label{eqn:chordless_factor}\\
    \boxtimes_n&=p^n\frac{N!}{n!(N-n)!} \text{.}
\end{align}
The last expression is obtained by calculating the number of potential ways in which $n$ vertices can be selected from a graph, which will have edges forming a cycle containing them with probability $p^n$.
Even though $\boxtimes_n$ gets combinatorially larger with increasing $N$ and $p$, the effect of increasing link probability is to reduce the number of those cycles that are chordless, making longer chordless cycles increasingly rare compared to shorter ones.
We plot in \Cref{fig:chordless_cycles} the theoretical count of chordless cycles in a random graph of $N=100$ nodes, for a range of link probabilities from $0.05$ to $0.85$.
The lower number is chosen to be above the critical value at which a fully connected graph emerges, given by the threshold $p \geq \frac{\log(N)}{N}$ \cite{barabasi2016network}.
Even for a graph of this size, the number of cycles becomes very large as $p$ increases, but for our graphs $p$ is much closer to $0.05$, at which point the most frequent chordless cycle is a square with a count of $85$.

Regardless of the rarity of long chordless cycles, it is well established in lattice field theory that ascending orders of a derivative of any quantity defined at a vertex, requires the consideration of vertices that are of increasing edge distance away \cite{creutz1983quarks}.
We make the simplifying assumption to restrict our analysis to a maximum cycle length of $n=5$, which captures vertices that are neighbors and next to nearest neighbors of a given vertex.
Considering the discrete calculus of a quantity defined at a vertex, this would permit the computation of a second derivative, which we consider sufficient for our analysis.
In principle any order of derivative could be included by extending the length of cycles included in our model.

The last simplifying assumption is the Independent Short Cycle (ISC) condition, which is the analog of the elastic particle assumption in the kinetic theory of gases, which we reproduce from \cite{tee2021enhanced} extended to cycles of any length.

\begin{definition}{The independent short-cycle condition.}\label{def:hard_core}
The independent short-cycle condition is satisfied by a graph $G(V,E)$ on $N$ vertices, if all of its closed cycles of length $n$ do not share more than one edge. Let $\square_n (e_{ij})$ represent a closed cycle supported upon an edge $e_{ij}$ of length $n$, with $n \geq 3$.  
The independent short-cycle condition is satisfied for the graph $G$ if and only if,
\begin{equation}
    \bigcap\limits_{n=3}^{n \leq N} \square_n (e_{ij}) = e_{ij} \text{,~} \forall e_{ij} \in E \text{.}
\end{equation}
\end{definition}

With these conditions we can state the mean-field results from \cite{tee2021enhanced} for the discrete curvatures $\kappa^f$ and $\kappa^o$ in terms of the number of triangles $\triangle_{ij}$, squares $\square_{ij}$ and pentagons $\pentagon_{ij}$ incident upon an edge $e_{ij}$.
Denoting by $k$ the average degree of a node in the graph we have the following expressions for the curvatures,

\begin{align}
    \kappa^{f}_{ij}&=4-2k +3 \triangle_{ij} + 2 \square_{ij} + \pentagon_{ij} \text{,} \label{eqn:exact_fr_expansion}\\
    k \kappa^{o}_{ij} &= 4-2k +3 \triangle_{ij} + 2\square_{ij} +\pentagon_{ij}+ \delta_{ij} \label{eqn:exact_or_expansion} \text{.}
\end{align}
where the correction to the mean field value $\delta_{ij}$ is given by Eq. \eqref{eqn:mf_cases}.

\begin{equation}\label{eqn:mf_cases}
\delta_{ij} = \begin{cases}
0 &\text{if $k > 2+ \triangle_{ij}+\square_{ij}+\pentagon_{ij}$}\\
k-2-\triangle_{ij}-\square_{ij}-\pentagon_{ij} &\text{if $2+\triangle_{ij}+\square_{ij} < k < 2+ \triangle_{ij}+\square_{ij} + \pentagon_{ij}$} \\
2k-4-2\triangle_{ij}-2\square_{ij}-\pentagon_{ij} &\text{if $ k < 2+\triangle_{ij}+\square_{ij}$}
\end{cases}
\end{equation}

In the case of FR curvature we can relax the ISC condition in the mean-field approximation somewhat, as the formula for FR curvature in \Cref{eqn:exact_fr_expansion} only relies upon it to compute the number of parallel edges, using the assumption that each cycle incident upon an edge is independent.
This enters in the equation when we compute the number of parallel edges incident upon the $0$-cells (i.e. vertices) $i$ and $j$ of an edge $e_{ij}$, but not part of any cycle  on $e_{ij}$, which we write as $P_0$.
We achieve this by subtracting from the average degree of the nodes at either end of an edge, two edges for each cycle incident upon the edge.
Specifically, for a mean field edge, the number of such parallel edges is $2k-2 - 2(\triangle_{ij}+\square_{ij}+\pentagon_{ij})$, on the assumption that each of the cycles are independent.
Therefore if we know the probability of independent edges we have an opportunity to further refine \Cref{eqn:exact_fr_expansion}.

Let us denote by $\rho_{ij}=(\triangle_{ij}+\square_{ij}+\pentagon_{ij})$ the sum of all cycles to order $5$ incident upon the edge $e_{ij}$, and by $\xi$ the probability that {\sl all} the edges $\rho_{ij}$ are independent according to \Cref{def:hard_core}.
The number of edges consumed by cycles is at least $2\Theta[(1-\xi)\rho_{ij}]$, where $\Theta[\dots]$ is the Heaviside step function with $\Theta[ 0 ]=0$.
The factor of $(1-\xi)$ is present to ensure that as $\xi \rightarrow 1$ we reflect that all contributions to $P_0$ comes uniquely from the cycles incident upon the edge and will be accounted for by $\rho_{ij}$.
Bringing this together, we estimate the number of parallel edges $P_0$ that arise from edges incident upon the vertices $i$ and $j$ as, 
\begin{equation}
    P_0=2k-2-2\Theta[(1-\xi)\rho_{ij}] -2\xi\rho_{ij} \text{.}
\end{equation}
The probability $\xi$ is another mean field parameter of the curvature, and will simplify our computations later.
It should be noted that our estimate of $P_0$ is an upper bound, as some non-independent cycles will still consume edges from the $2k-2$ available ones at the vertices $i$ and $j$.

If we make the assumption that $\rho_{ij} > 0$, which for the ground states obtained for IG models is a reasonable one, we arrive at the following modified version of \Cref{eqn:exact_fr_expansion} that we will find useful in our numerical simulations,
\begin{equation}\label{eqn:non_isc_fr}
    \kappa^f_{ij} = 4-2k + 2\Theta[(1-\xi)\rho_{ij}] + (1+2\xi)\triangle_{ij} + 2\xi \square_{ij} - (2\xi-1)\pentagon_{ij} \text{.} 
\end{equation}
A similar expression is not available for OR curvature.

The parameter $\xi$ can be estimated by computing the number of ISC violating edges in the graph.
In Kelly {\sl et al} \cite{kelly2019self} there is a simple method for identifying edges that violate the ISC constraint that rests upon searching the graph for certain motifs, a technique we make use of in our simulations.
If $|E|_{v}$ is the number of edges that violate the ISC constraint, then
\begin{equation*}
    p_{isc}=1-\frac{|E|_{v}}{|E|} \text{,}
\end{equation*}
is the probability that an edge satisfies it, and we will make use of the following estimate for $\xi$, obtained from the average value of $\rho_{ij}$,
\begin{equation}
    \xi=p_{isc}^{\expval{\rho_{ij}}} \text{.}
\end{equation}

Although mean field, the result in \Cref{eqn:non_isc_fr} is still exact with respect to the cycles.
As we are intending to use the scalar values for the discrete curvature we can further simplify by computing the average values of $\triangle_{ij}, \square_{ij}$ and $\pentagon_{ij}$ in terms of powers the adjacency matrix, which we can obtain using the cycle count formulae originally derived by Harary {\sl et al} and Perepechko {\sl et al } \cite{harary1971number,perepechko2009number}.
If we denote $\triangle$, $\square$, and $\pentagon$ as the total number of $3,4$ and $5$ cycles in the graph, we state the following result due to Harary  \cite{harary1971number},

\begin{align}
    \triangle &= \frac{1}{6} \Tr ( A^3) \label{eqn:triangles} \mbox{,}\\
    \square  &= \frac{1}{8} \Tr (A^4) - \frac{\abs{E}}{4} - \frac{1}{4} \sum\limits_{i \neq j} \sum\limits_{k \neq i,j} A_{ik} A_{kj} \label{eqn:squares} \mbox{,} \\
    \pentagon &= \frac{1}{10} \Tr (A^5) - \frac{1}{2}\Tr (A^3) -\frac{1}{2} \sum\limits_i \left ( \sum\limits_j A_{ij}-2 \right ) A^3_{ii} \label{eqn:pentagons} \mbox{.}
\end{align}

Using these expressions for the total number of cycles in a graph with $\abs{E}$ edges, we can compute the average number of a given cycle incident upon a given edge by assuming that any edge of an $n$ polygon has probability $n/\abs{E}$ of containing that edge.
For each polygon we obtain,
\begin{equation}\label{eqn:mean_cycle}
    \expval{\triangle_{ij}} = \frac{3 \triangle}{\abs{E}} \mbox{,}
    \expval{\square_{ij}} = \frac{4 \square}{\abs{E}} \mbox{,}
    \expval{ \pentagon{ij}} = \frac{5 \pentagon}{\abs{E}} \mbox{,}
\end{equation}
which we can substitute into Eqs. \eqref{eqn:exact_fr_expansion} and \eqref{eqn:exact_or_expansion} to obtain values of $\expval{\kappa^f_{ij}}$ and $\expval{\kappa^o_{ij}}$.
We can then compute $\expval{\kappa^f}$ and $\expval{\kappa^o}$ by recalling that $\expval{\kappa^f} = 2|E| \expval{\kappa^f_{ij}}$ with the parallel result for OR curvature.

It should be noted that Eqs. \eqref{eqn:triangles}, \eqref{eqn:squares} and \eqref{eqn:pentagons} all refer to the total number of cycles i.e. $\boxtimes_n$, not $\square_n$ and so when we come to use them in our simulations we will have to adjust using Eq. \eqref{eqn:chordless_factor}.

Before stating the expressions for curvature obtained by substituting in these mean field expressions for cycles into those for the curvature, it is instructive to examine the extremal properties of the Hamiltonians we obtain when using the prescription in Eq. \eqref{eqn:h_comb}.
If we again take $p$ to indicate the probability of an edge between two randomly chosen vertices, we note that for an average edge, after traversing the $k-1$ edges connected to one end of the edge to close cycles to the other end of the edge we have,
\begin{align}
    \expval{\triangle_{ij}}  &= 2(k-1) p \text{,}\\
    \expval{\square_{ij}}    &= 2(k-1) p^2 \text{,} \label{eqn:exp_squares}\\
    \expval{\pentagon_{ij}}  &= 2(k-1) p^3 \text{.}
\end{align}
In the ground states of our models for a graph of size $N$ we find typically that $p \ll 1$, and indeed formally $p \rightarrow 0$ as $N \rightarrow \infty$, and so we can write our curvatures in the sparse graph limit of very small $p$ as,
\begin{align*}
    \expval{\kappa^f} &=N k( 4-2k) + \mathcal{O}(p) \text{,} \\
    \expval{\kappa^o} &=N( 4-2k) + \mathcal{O}(p) \text{.}
\end{align*}
We have made the assumption that the term in $2\Theta[(1-\xi)\rho_{ij}]$, does not contribute in the sparse graph approximation as in the limit $p \rightarrow 0$ we have $\expval{\rho_{ij}}=0$.

Similarly for the kinetic term in the Hamiltonian we will pick up a contribution of $Nk$, assuming that all connected spins are aligned (which is precisely the form of ground state encountered in \cite{trugenberger2015quantum,tee2020dynamics}).
With this assumption we obtain the following expressions for the Hamiltonian as a function of $k$ for both curvature measures as,
\begin{align}
    H^f(k) &= \frac{g^2}{4} Nk(2k-4) -\frac{g}{2}Nk + \mathcal{O}(p) \text{,} \label{eqn:fr_hamiltonian}\\
    H^o(k) &= \frac{g^2}{4} N(2k-4) -\frac{g}{2}Nk + \mathcal{O}(p) \label{eqn:or_hamiltonian} \text{, }
\end{align}
noting that $\expval{\kappa^f}=2|E| \expval{\kappa^f_{ij}} = Nk\expval{\kappa^f_{ij}}$.
We immediately see that there is no well defined minimum of the energy for the OR curvature case by virtue of the missing factor of $k$.
It is entirely possible, however, that higher order of $p$ contributions will bring in terms of at least $k^3$ that would allow a well defined minimum of \Cref{eqn:or_hamiltonian}.
Unfortunately in our simulations all attempts to use OR curvature to find stable ground states for the CIM models has proven elusive, and we will henceforth concentrate on the models utilizing FR curvature.

It should be stressed that the approximation used to argue the existence of a minimum for the Hamiltonian constructed from FR curvature, $H^f (k)$, is only valid for small link probability, which although a safe approximation as we shall see in \Cref{sec:results}, does not reproduce the zero curvature for square lattices.
Putting aside temporarily this complication, we can identify the minimum of \Cref{eqn:fr_hamiltonian} as a function of $g$ using elementary calculus.
The minimum of $k$, $k_m$ is obtained as $k_m=\frac{1}{2g} + \frac{1}{2}$.
In a $d$ dimensional regular lattice $k=2d$, and so the dimension of such a  ground state graph corresponding to the minimum of \Cref{eqn:fr_hamiltonian}, which we denote by $d_k$, has the value,
\begin{equation}\label{eqn:naive_dimension}
    d_k=\frac{1}{4g} + \frac{1}{2} \text{.}
\end{equation}
If we extend the approximation to $\mathcal{O}(p^2)$ to include squares, this expression becomes  more complex.
Specifically our expression for the curvature becomes $\expval{\kappa_{ij}}=4-2k +6(k-1)p + 4(k-1)p^2$.
If we define $\epsilon=3p+2p^2$, we can minimize the resultant expression for the Hamiltonian and obtain,

\begin{equation}
    d_k = \frac{1}{1-\epsilon} \left \{ \frac{1}{4g} + \frac{1}{2} - \frac{\epsilon}{4} \right \} \text{.}
\end{equation}
As $\epsilon \rightarrow 0$ we recover the original expression \Cref{eqn:naive_dimension}, and it is instructive to consider typically how small $\epsilon$ is in the graphs we will encounter as ground states in \Cref{sec:results}.
For a graph of size $N=100$, which as we shall see becomes a highly regular lattice at around $k=6$, $\epsilon=0.185$.
Our dimension $d_k$ will be approximately $27\%$ bigger when we include terms to order $\mathcal{O}(p^2)$.
If we repeat that calculation for $N=1000$, this drops to a $0.45\%$ increase in the  dimension.
Given that the assumption for our model of spacetime is that the ground state graphs are extremely large, we will accept the approximation in \Cref{eqn:naive_dimension}, and we will refer to $d_k$ as the `naive dimension' of the graph.

\subsection{The discrete canonical Hamiltonian and comparison with the Ising models}
\label{ssec:discrete_hamiltonian}
To conclude the theoretical investigation of the CIM model let us capitalize upon the relations due to Harary in \Cref{eqn:triangles}, \Cref{eqn:squares} and \Cref{eqn:pentagons} to obtain an expression for the Hamiltonian in terms of the graph adjacency matrix.

First of all we note that by substituting in the expressions for the expected value of the short cycles stated in \Cref{eqn:mean_cycle} into our ISC corrected formula for the FR curvature of an edge \Cref{eqn:non_isc_fr}, we obtain,
\begin{equation}
\begin{split}
    \expval{\kappa^f_{ij}}&=4-2k + 2\Theta[(1-\xi)\rho_{ij}] \\
    & + \frac{1}{|E|}\left \{ 3(1+2\xi)\triangle + 8\xi \square + 5(2\xi-1)\pentagon  \right \}\text{.}
\end{split}
\end{equation}

To obtain the expected FR scalar we multiply through again by $2|E|$ yielding,
\begin{equation}
\begin{split}
    \expval{\kappa^f}&=2|E|(4 + 2\Theta[(1-\xi)\rho_{ij}] -2k) \\
    & +6(1+2\xi)\triangle + 16\xi \square + 10(2\xi-1) \pentagon \text{,}
\end{split}
\end{equation}
which is the form we shall use in our simulations.
For completeness we can insert this directly into the expression for the $H_{CIM}$ Hamiltonian in \Cref{eqn:h_comb}, to obtain,
\begin{equation}\label{eqn:h_cim_ulation}
\begin{split}
    H_{CIM}=&-\frac{g^2}{4} \left \{  2|E|\Big\{4 + 2\Theta[(1-\xi)\rho_{ij}] -2k\Big\}+6(1+2\xi)\triangle + 16\xi \square + 10(2\xi-1) \pentagon \right \} \\ &-\frac{g}{2}\sum\limits_i \sum\limits_j \hat{s}_i A_{ij} \hat{s}_j \text{.}
\end{split}
\end{equation}
In the simulations we will directly compute this energy function for the graph, but it is instructive to go one step further and rewrite this expression in terms of the adjacency matrix of the graph.
This is accomplished by using Equations, \eqref{eqn:triangles}, \eqref{eqn:squares} and \eqref{eqn:pentagons}.
However we must also take into account the fact that not all pentagons and squares are chordless, and so we need to factor down these expressions using \Cref{eqn:chordless_factor}.
For a link probability of $p$, these factors are $p_4=(1-p)^2$ for a square, and $p_5=(1-p)^3$ for a pentagon.
After some algebra, substituting these factored expressions for graph cycles into \Cref{eqn:h_cim_ulation}, we obtain  the following form of our Hamiltonian,

\begin{equation}\label{eqn:h_cim_adjacency_full}
\begin{split}
    H_{CIM}&=\frac{g^2}{2} \Bigg \{ \left[ (5p_5-1)\xi - \frac{1}{2}(1+5p_5)\right ]\Tr (A^3)  + 2p_4\xi \sum\limits_{i\neq j}\sum\limits_{k \neq i,j} A_{ik}A_{kj}\\
    &+ \frac{5}{2}p_5(2\xi-1) \sum\limits_i \left ( \sum\limits_j A_{ij}-2 \right ) A^3_{ii} +2|E|\Big \{k+ p_4 \xi - 2 - \Theta[(1-\xi)\rho_{ij}]\Big\}\\
    &-p_4 \xi \Tr (A^4) - \frac{1}{2}p_5(2\xi-1) \Tr (A^5)   \Bigg \} -\frac{g}{2}\sum\limits_i \sum\limits_j \hat{s}_i A_{ij} \hat{s}_j \text{.}
\end{split}
\end{equation}
This expression is complex, but necessary for the simulations that we describe in the next section.
Consider the case with $p_4=p_5 \approxeq 1.0$ and $\xi=1$, which corresponds to the case of an ideal graph with the ISC fully satisfied and most cycles chordless.
As $2\Theta[(1-\xi)\rho_{ij}]$ is zero for $\xi=1$, \Cref{eqn:h_cim_adjacency_full}  simplifies to,
\begin{equation}\label{eqn:h_cim_adjacency}
\begin{split}
    H_{CIM}&=\frac{g^2}{2} \Bigg \{ \Tr (A^3)  + 2 \sum\limits_{i\neq j}\sum\limits_{k \neq i,j} A_{ik}A_{kj} + \frac{5}{2} \sum\limits_i \left ( \sum\limits_j A_{ij}-2 \right ) A^3_{ii}\\
    & +2|E|(k-1) - \Tr (A^4) - \frac{1}{2} \Tr (A^5)  \Bigg \} -\frac{g}{2}\sum\limits_i \sum\limits_j \hat{s}_i A_{ij} \hat{s}_j \text{.}
\end{split}
\end{equation}
By comparing this expression for the Hamiltonian of CIM expanded in terms of the adjacency matrix and the Hamiltonian for QMD in \Cref{eqn:qmd_hamiltonian}, we note a similarity in form.
In particular, the leading two terms in $\frac{g^2}{2}\{\}$ bracket are both present in the QMD Hamiltonian, as is the last term that expresses the kinetic contribution.
However, in the $H_{CIM}$ Hamiltonian, the $\sum A_{ik}A_{kj}$ has a factor of $2$.
In fact the $\sum A_{ik}A_{kj}$ expression becomes less significant than the $2|E|(k-1)$ term due to the fact that $\xi$ is typically very small, but fortunately minimizing the $2|E|(k-1)$ term is equivalent to minimizing $\sum A_{ik}A_{kj}$.
To see this, let us consider the mean field approximation of $2|E|(k-1)$.
For a given node $i$ of average degree $k$,  $\sum\limits_{mn} A_{mi}A_{in}$  counts the number of edge pairs, or open triples, with node $i$ being the central node.
In the mean field approximation there are $\frac{1}{2}k(k-1)$ for each vertex.
As the summation is over $i$ and $j$, each vertex is counted twice, and so,
\begin{equation*}
    \sum\limits_{i\neq j}\sum\limits_{k \neq i,j} A_{ik}A_{kj} = Nk(k-1)=2|E|(k-1) \text{,}
\end{equation*}
as $Nk=2|E|$.
Going back to \Cref{eqn:h_cim_adjacency_full}, we can make this substitution for the $2|E|(k-1)$ term and write out explicitly the relationship between $H_{CIM}$ and $H_{QMD}$ as,
\begin{equation}\label{eqn:qmd_cim}
\begin{split}
    H_{CIM}&=\frac{g^2}{2} \Bigg \{ \Tr (A^3)  + \sum\limits_{i\neq j}\sum\limits_{k \neq i,j} A_{ik}A_{kj} + \dots  \Bigg \} \\ &-\frac{g}{2}\sum\limits_i \sum\limits_j \hat{s}_i A_{ij} \hat{s}_j\\
    &+\frac{g^2}{2} \Bigg \{ [(5p_5-1)\xi -\frac{1}{2}(3+5p_5)] \Tr (A^3) \\
    &+ 2p_4\xi \sum\limits_{i\neq j}\sum\limits_{k \neq i,j} A_{ik}A_{kj} \Bigg\} \\
    &= H_{QMD} \\
    &+ f(p_5,\xi) \Tr (A^3) + 2p_4\xi \sum\limits_{i\neq j}\sum\limits_{k \neq i,j} A_{ik}A_{kj} + \dots\text{,}\\
    & \text{where~} f(p_5,\xi)=[(5p_5-1)\xi -\frac{1}{2}(3+5p_5)] \text{.}
\end{split}
\end{equation}
The $``\dots"$ in \Cref{eqn:qmd_cim} refers to the  omitted  terms in powers of $A_{ij}$ higher than $3$. 
The similarity to the QMD Hamiltonian is now clear.
Indeed the additional terms are all dependent upon $p_4,p_5$ and $\xi$, which vary as the structure of the graph changes, specifically the total number of edges.
In the limit of zero connectivity $p_4=p_5=\xi=1.0$, at which point $H_{QMD}=H_{CIM}$, but this is of course a trivial scenario.
For low values of $d_k$, corresponding to $p_4$ and $p_5$ approaching $1.0$,  the behavior is complex.
Providing $\xi < 0.5$, the complete $\Tr (A^3)$ term, including the additional corrections, will change sign  in $H_{CIM}$ and start favoring triangles as $d_k$ reduces.
This will be  somewhat offset by an increase coefficient in the $\sum\limits_{i\neq j}\sum\limits_{k \neq i,j} A_{ik}A_{kj}$ term that disfavors open triples in the graph.
Indeed we note that $f(p_5,\xi)$ is bounded in the range $[-4,0]$, and that when $f(p_5,\xi) < -1$ the $\Tr (A^3)$ contribution will have an overall negative coefficient.
Accordingly we expect our CIM model to share many feature with QMD, but not yield exactly equivalent ground state graphs.
In \Cref{sec:results} we see precisely this behavior with ground state graphs for the two models that are quite similar.
The $H_{CIM}$ ground state, however, exhibits a reemergence of triangles and clustering for low values of $d_k$.
This would be consistent with sign of the $\Tr (A^3)$ term reversing at low dimensions.

We believe this analysis justifies our interpretation of QMD as an approximation to the full CIM model.
The additional powers of the adjacency matrix that come from considerations of higher order cycles, represent corrections to the estimate of the discrete curvature, supporting the intuition that all the IG Hamiltonians are an approximation to a more physical model of emergent discrete geometry.
We will now proceed to explore the extent of this similarity by computing numerically the ground states of the CIM and QMD model in order to compare the structure of the obtained ground state graphs.

\section{Simulation results and discussion}
\label{sec:results}
\subsection{Computing the ground state}
Using the results for the Hamiltonian derived in \Cref{ssec:discrete_hamiltonian}, we can compute the ground state graphs of fixed size, varying the model's free parameter, $g$, the coupling constant.
We use the same technique of Glauber dynamics previously described in \cite{trugenberger2015quantum, tee2020dynamics}, which itself is adapted from techniques commonly used in certain classes of neural networks \cite{muller2012neural}.
The additional terms in powers of the adjacency matrix present in the Hamiltonian $H_{CIM}$, require a slight modification to accommodate them.
We adapt the method used for the QMD simulations described in \cite{tee2020dynamics}, to account for the $\Tr A^3_{ij}$ term present in the QMD model and also in the extensions to DGM described in \cite{trugenberger2016random}.
To summarize we start with a fully randomized graph with $N$ vertices, where the orientation of the vertex spins is set $\ket{1}$ or $\ket{0}$ with equal probability, and edges set to exist between all pairs of vertices with probability $p=0.5$.
Sequentially a sample of the vertices and edges is chosen.
With even probability we change the spin states, retaining the change if it lowers the energy by being aligned to more of a vertices neighbors.
If the energy is not reduced we restore the previous spin state.
We then address the sample of edges, again randomly adding or removing edges between the vertices in the sample set.
For each change in the edge set we compute an edge energy function that estimates the effect the change will have on the overall Hamiltonian of the model.
If the random change reduces the value of the energy function the change is accepted. 
The whole change set is then accepted or rejected if the total graph Hamiltonian is reduced.
By iterating this approach a configuration of the graph that minimizes the Hamiltonian is achieved, and the simulation is averaged over multiple runs to restrict the influence of any individual run reflecting a non global minimum and distorting the results.

In detail, the process of minimization using Glauber dynamics starts with a random sample of the vertices, and we alter the value of the spin state with even probability to the opposite of the initial value.
At each time step, for each vertex we compute,
\begin{equation}\label{eqn:spin_itr}
	 h_i=\sum\limits_{j \in V } A_{ij} s_j \mbox{,}
\end{equation}
and if $h_i \geq 0$ set $s_i=+1$, otherwise $s_i=-1$.
We now select a subset of vertex pairs $i$ and $j$, which may or may not have an edge between them.

For each of the potential edges $e_{ij}$ in our selected sample, we now randomly alter the edge state to the {\sl opposite} of what it was, recording the change in a variable $\delta$, which we set to $+1$ if we add an edge and $-1$ for removal.
We now seek to calculate the effect of the change on the total Hamiltonian by defining a measure of the  edge energy difference $h_{ij}$ for a change of edge state.
To compute the contribution of this edge to the Hamiltonian, we make the approximation that for terms involving a power of the adjacency matrix such as $A^n_{ij}$ the contribution can be obtained by comparing the change in the trace of the $n^{th}$ power of the matrix at the vertices $i$ and $j$.
The addition or removal of an edge $e_{ij}$, will potentially change both $A^n_{ii}$ and $A^n_{jj}$, and so we average these changes to determine the contribution to  $h_{ij}$.
Wherever terms such as  $\Tr(A^n)$ appear in the Hamiltonian we substitute $\Delta(A^n)= \frac{1}{2}\{\Delta(A^n_{ii}) + \Delta(A^n_{jj})\}$ in the expression for the edge energy function $h_{ij}$.
For the $\sum A_{ik}A_{kj}$ term, an edge will contribute $\delta( k_i + k_j -1)$ to the edge energy difference, with $\delta$ recording the direction of the change.
Turning to the $2|E|\Big \{k+ p_4 \xi - 2 - \Theta[(1-\xi)\rho_{ij}]\Big\}$ term, which as we discussed in the prior section, is in the mean field approximation, equivalent to an additional $\sum A_{ik}A_{kj}$ term.
For the change in edge state $k$ changes by $\delta$, and the contribution to the $h_{ij}$ arises from this variation in $k$, giving $\delta 2|E|$.
As we are considering a single edge, we factor by the change to $|E|$ from both vertices $i$ and $j$, which is $\delta(k_i + k_j)$, recalling that $\sum k=2|E|$.
Bringing all of these together and noting that the contribution must be doubled as each edge is counted twice in $H_{CIM}$, we arrive at our final edge energy function below,
\begin{equation}
\begin{split}\label{eqn:energy_itr}
	 	h_{ij}&= g^2  \Bigg \{  \left[ (5p_5-1)\xi - \frac{1}{2}(1+5p_5)\right ]\Delta(A^3)  -2p_4\xi\Delta(A^4) \\
	 	& + \delta(k_i+k_j) + 2\delta p_4\xi(k_i+k_j-2) - p_5(2\xi-1)\Delta(A^5)\\ 
	 	& + 5p_5(2\xi-1) \sum\limits_{ij} (A_{ij}-2 ) \Delta(A^3) \Bigg \} \\
			&-\delta g s_i s_j  \text{.}
\end{split}
\end{equation}
If $h_{ij} \leq 0$ we retain the change, and if not it is discarded.
At the end of the evaluation of every link in the sample we compute the value of the Hamiltonian \Cref{eqn:h_cim_ulation}.
If the Hamiltonian has reduced we retain the change set, otherwise we discard all of them.
Eventually we will arrive at a stable minimum, which we define as successive iterations not discovering a configuration of overall lower energy.
As we approach the minimum, we reduce the sample size of nodes and edges,  and, lengthen the number of successive iterations without change required to indicate that the model has minimized its energy.
This allows us to finer grain the calculation of the ground state graph configuration.
For all ground states we repeat the calculation $10$ times and average the harvested metrics across each of the runs.
The choice of $10$ runs is dictated by computational time, with each computation of the ground state at $N=100$ nodes requiring $20$ to $30$ minutes to complete on mid-size server hardware.
Our results could be significantly improved by the use of more powerful hardware.

In addition to the computation of the ground states we also seek to investigate the emergence of a preferred low valued dimension, a feature of the QMD and DGM models.
We  repeat the analysis conducted in earlier work, using the method described above for computing the ground state graphs of the CIM and QMD Hamiltonian.
We focus on the Hausdorff measure of extrinsic dimension, Spectral measure for intrinsic dimension \cite{tee2020dynamics,ambjorn1997quantum}, the `lattice' dimension described below, and finally the `naive dimension' $d_k$ described by \Cref{eqn:naive_dimension}.
To compute these dimensions we take a slightly different approach to that taken in \cite{tee2020dynamics,trugenberger2015quantum}, instead focusing on a smaller number of relatively larger graphs of $N=350$ nodes obtained by computing the ground state of our Hamiltonians.
Randomization of the results is obtained by repeated computations of the dimensions seeded from different starting nodes in the graphs.

\begin{figure}[htbp]
	\centering
	\includegraphics[scale=0.43]{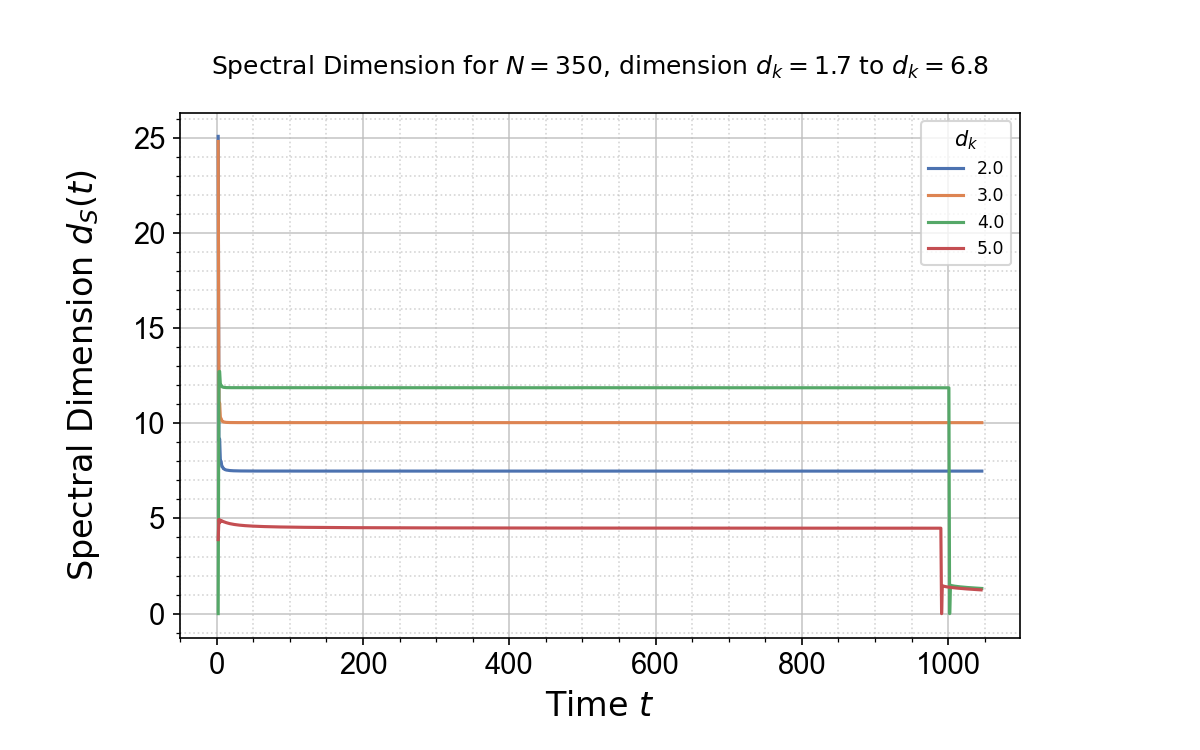}
	\caption{We compute and plot the spectral dimension $d_S$ as a function of time $t$ for a graph of $N=350$ nodes. These are computed for a range of coupling constants $g$, represented by the equivalent naive dimension $d_k$ computed using \Cref{eqn:naive_dimension}.}
	\label{fig:spectral}
\end{figure}

The spectral dimension $d_S$ is problematic, as the computation relies up the return probability $p_r$ of a random walk to a randomly chosen starting node after $t$ steps, which is known to vary as,
\begin{equation}\label{eqn:spectral_d}
    p_r=t^{\frac{-d_S}{2}}\text{.}
\end{equation}
This result is only truly valid in the limit of an infinite graph, $N \rightarrow \infty$ \cite{ambjorn1997quantum}.
To overcome this limitation we conduct a series of random walks, of increasing time steps $t$, and then attempt a least squares fit to \Cref{eqn:spectral_d}, extracting from the best fit the value of  $d_S$.
We plot the results of this computation in \Cref{fig:spectral} for the CIM model.
We can see that the spectral dimension establishes a plateau until the length of the walk exceeds a multiple of the size of the graph, which depends upon the value of $d_k$.
This demonstrates the unreliability of $d_S$ for large values of $t$, which as $t \rightarrow \infty$, by definition $p_r \rightarrow 1$, evidenced by a sudden drop and then decay of the dimension to zero as $t$ increases.
The `cliff' occurs at lower values of $t$ for higher values of $d_k$, which is due to a more clustered ground state graph.
The value of $d_S$ of the largest non-zero plateau is used as the spectral dimension of the graph.

Hausdorff dimension $d_H$ is simpler, and is extracted by examining how the graph volume scales with distance.
We compute this by selecting a series of random nodes and then computing the scaling of the number of nodes at increasing shortest path distance to it.
This equates the volume $V$ with the cardinality of the subgraph at $r$ hops from the randomly selected node, and exploits the relation \cite{ambjorn1997quantum}
\begin{equation}\label{eqn:hausdorff}
r \propto V^{1/d_H} \text{,}
\end{equation}
to extract the value of the Hausdorff dimension. 
The distance $r$ can easily be computed using the the powers of the adjacency matrix to create a shortest distance matrix between all pairs of nodes in the graph.
We progressively raise the adjacency matrix to powers of increasing $n$.
The smallest value of $n$ that has a non-zero value at $A^n_{ij}$ is the shortest distance between vertices $i$ and $j$.

Finally, we define the lattice dimension of the graph $d_l$ to be half the average node degree, that is $d_l=\frac{1}{2}\expval{k}$.
This is an exact result for a square lattice graph $\mathbb{Z}^{d_l}$, where it is equal numerically to both the extrinsic and intrinsic dimension of the graph.
We make use of lattice dimension in our numerical results to understand how the various measures of dimension compare to this idealized dimension.
In particular, the closer the naive dimension is to the lattice dimension, the more the graph will resemble a regular square lattice.

We also compute for the ground state graphs the Von Neumann Entropy (VNE), which arises directly from the eigenvalue spectrum of the matrices representing the graph structure.
In information theory it is possible to quantify the information in `bits' required to describe the structure of a graph, and from that construct an entropy measure similar to Shannon entropy \cite{shannon1948mathematical}.
The VNE entropy for a graph was proposed as an analogy to the Von Neumann entropy of a quantum system defined in terms of its density matrix \cite{von2018mathematical}.
If one interprets the graph as the entanglement of the quantum states of the vertices and edges, it is natural to propose the Laplacian matrix of the graph as a density matrix \cite{passerini2008neumann,anand2009entropy}

Consequently the VNE entropy is computed by solving the eigenvalue problem for $L_{ij}$, obtaining the set of $N$ eigenvalues $\lambda_i$.
These are then used to define the dimensionless entropy,

\begin{equation}\label{eqn:vonneuman}
	S_G=-\sum\limits_{i=1}^{N} \lambda_i \log_2 \lambda_i \mbox{,}
\end{equation}
We compute this entropy for all of our simulations.

Finally for all of the ground state graphs we also compute degree distributions and the two-point degree correlation functions.
The two-point degree correlation is obtained by calculating the degree correlation matrix $e_{k k'}$ \cite{pastor2001dynamical,vazquez2002large,newman2003mixing}, which measures the probability of two randomly chosen nodes, connected by a single edge, having degree $k$ and $k'$.
From the degree correlation matrix we then compute the trace, which is the total probability of any randomly selected link connecting nodes of the same degree.
In a regular square lattice we would expect $\Tr e_{kk'}$ to be close to $1.0$, indicating a highly assortative graph where nodes tend to be connected to nodes of the same degree.

\subsection{Ground state metrics}

\begin{figure*}[htbp]
	\centering
	\begin{subfigure}[t]{0.45\textwidth}
		\centering
		\includegraphics[scale=0.39]{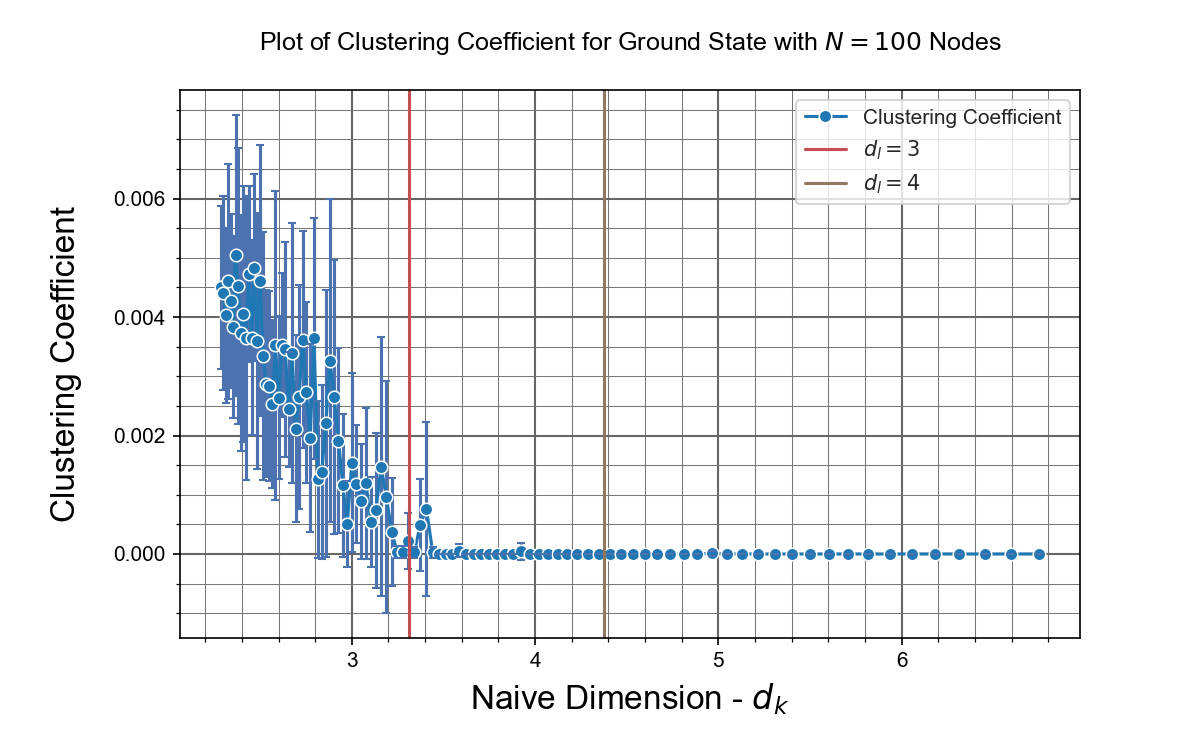}
		\caption{Averaged clustering coefficient, $N=100$.}
		\label{fig:CC100}
	\end{subfigure}%
	~ 
	\begin{subfigure}[t]{0.45\textwidth}
		\centering
		\includegraphics[scale=0.39]{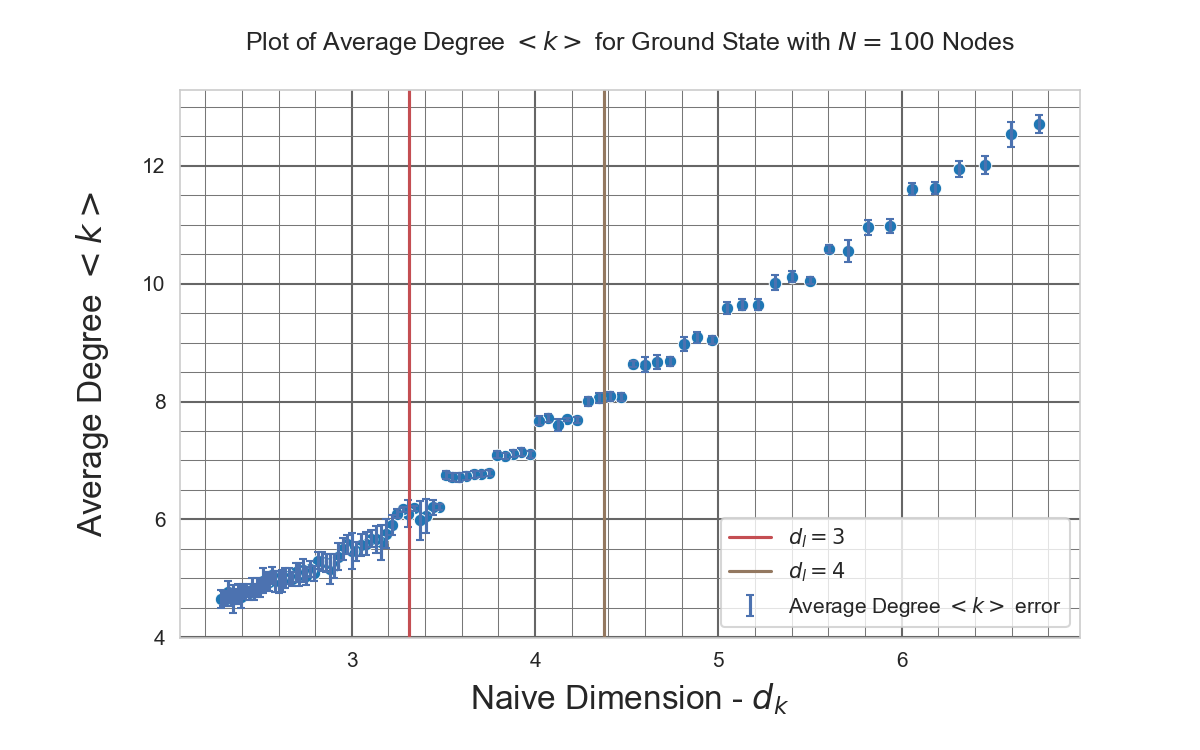}
		\caption{Averaged node degree, $N=100$.}
		\label{fig:Dgr100}
	\end{subfigure}
	~ 
	\begin{subfigure}[t]{0.45\textwidth}
		\centering
		\includegraphics[scale=0.39]{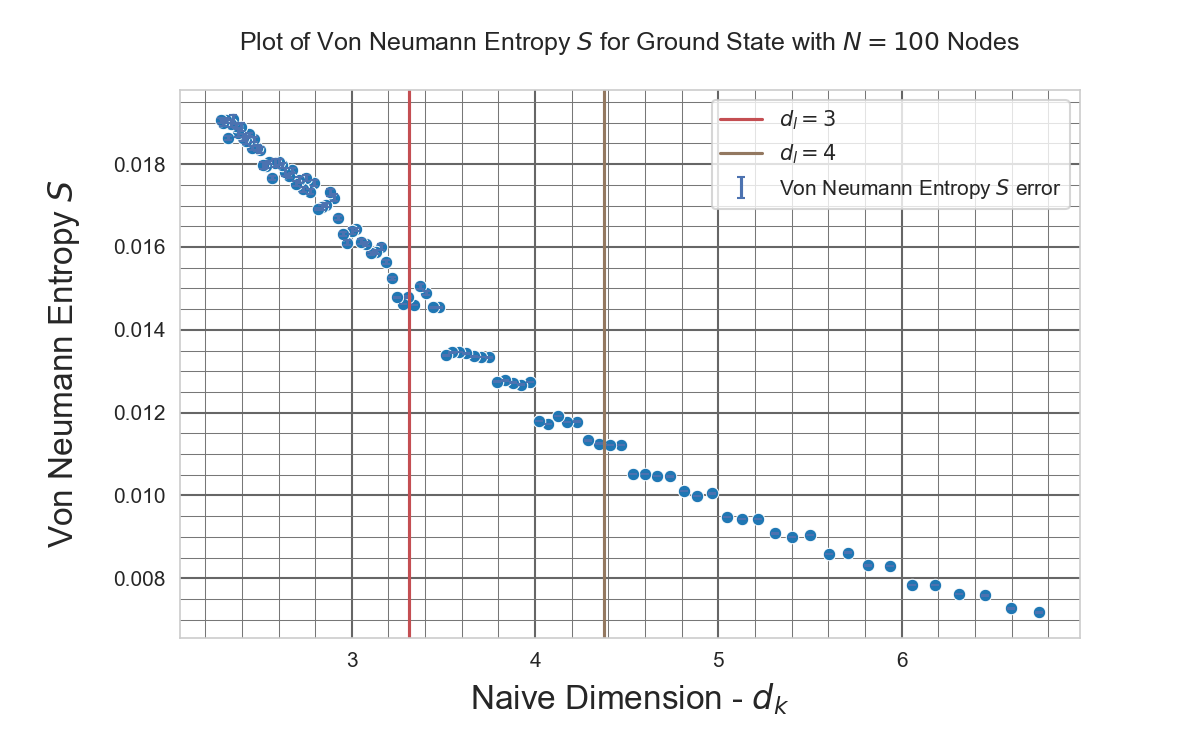}
		\caption{Averaged Von Neumann entropy density, $N=100$.}
		\label{fig:FR-VNE100}	
	\end{subfigure}%
	~ 
	\begin{subfigure}[t]{0.45\textwidth}
		\centering
		\includegraphics[scale=0.39]{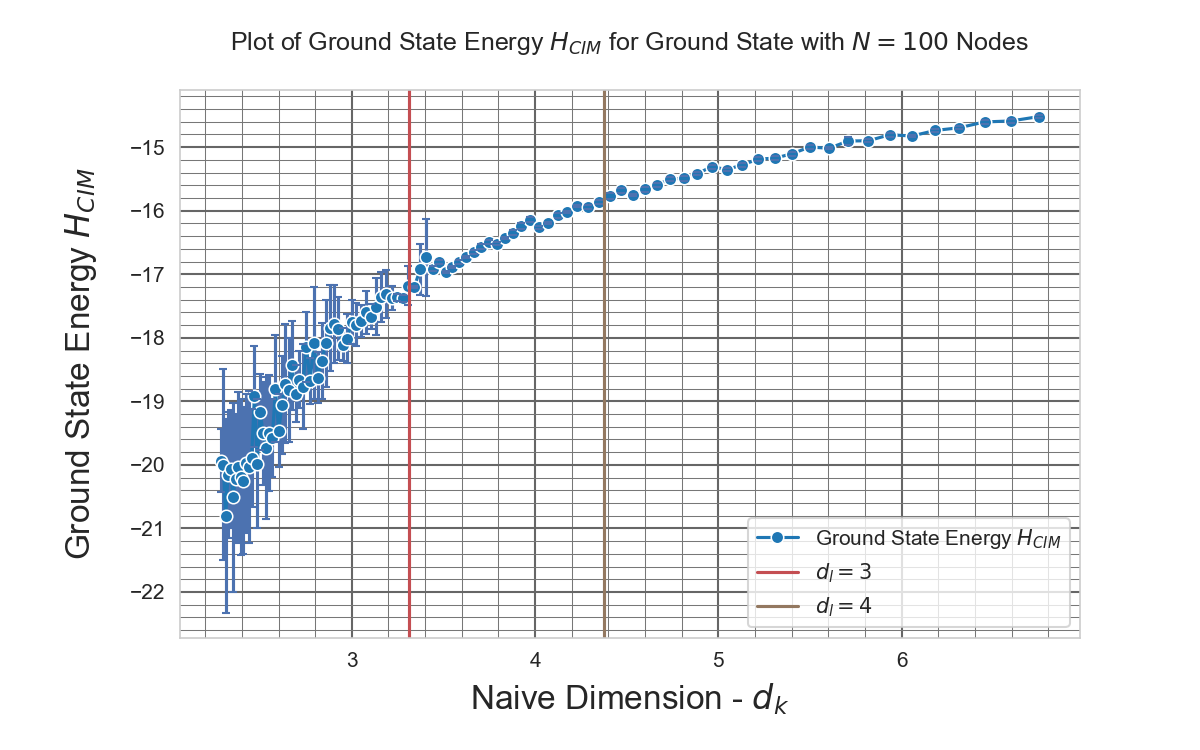}
		\caption{Averaged Graph Hamiltonian, $N=100$.}
		\label{fig:FR-H100}
	\end{subfigure}
	~
	\begin{subfigure}[t]{0.45\textwidth}
		\centering
		\includegraphics[scale=0.39]{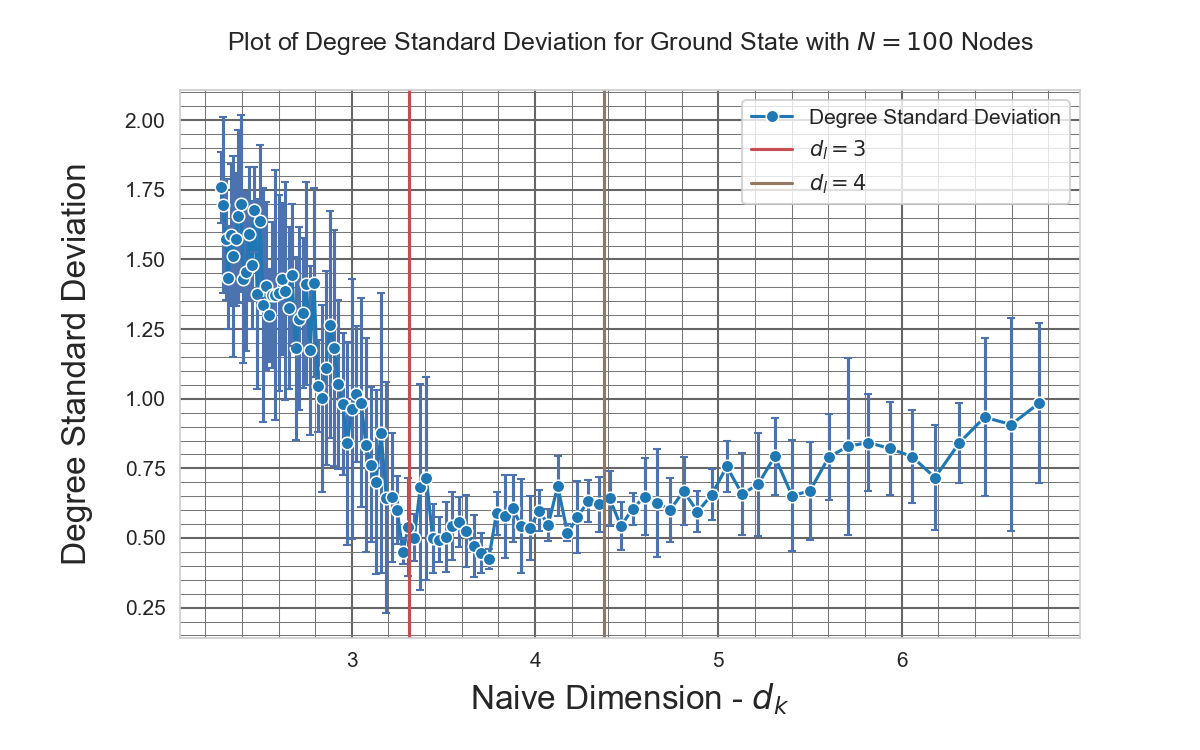}
		\caption{Standard deviation of average node degree, $N=100$.}
		\label{fig:kStd100}	
	\end{subfigure}%
	~ 
	\begin{subfigure}[t]{0.45\textwidth}
		\centering
		\includegraphics[scale=0.39]{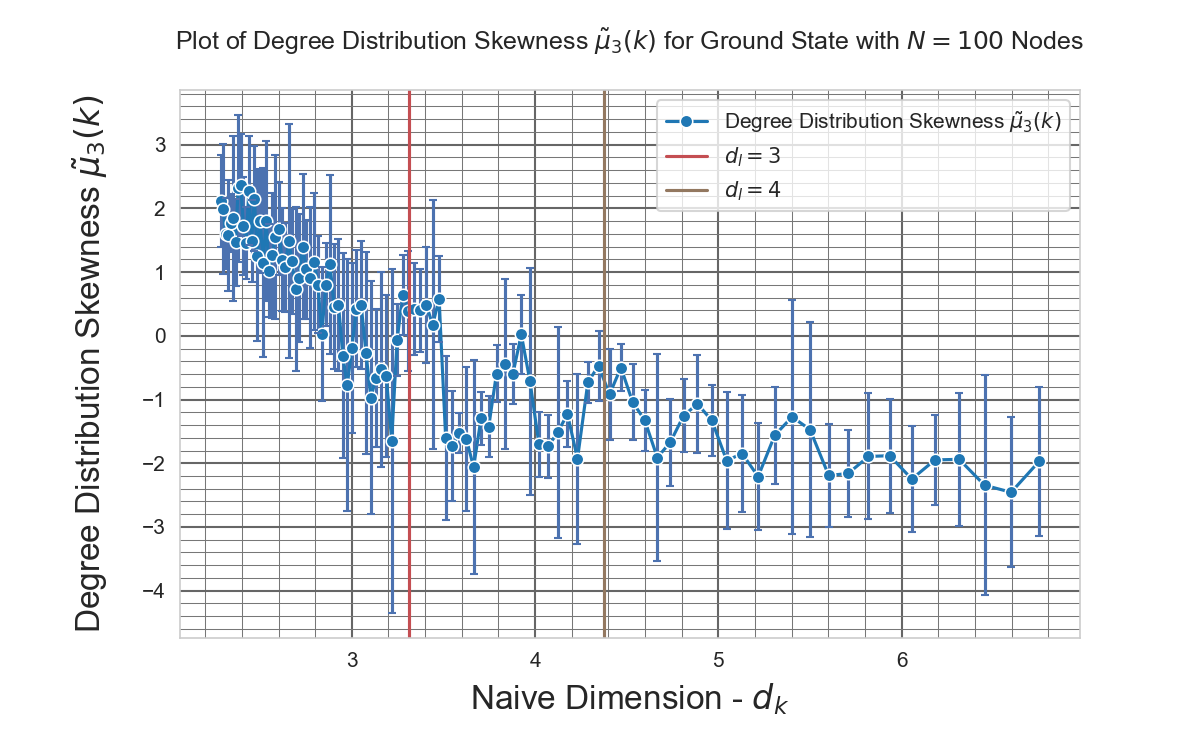}
		\caption{Skewedness of degree distribution, $N=100$.}
		\label{fig:kSkew100}
	\end{subfigure}
	~ 
	\begin{subfigure}[t]{0.45\textwidth}
		\centering
		\includegraphics[scale=0.39]{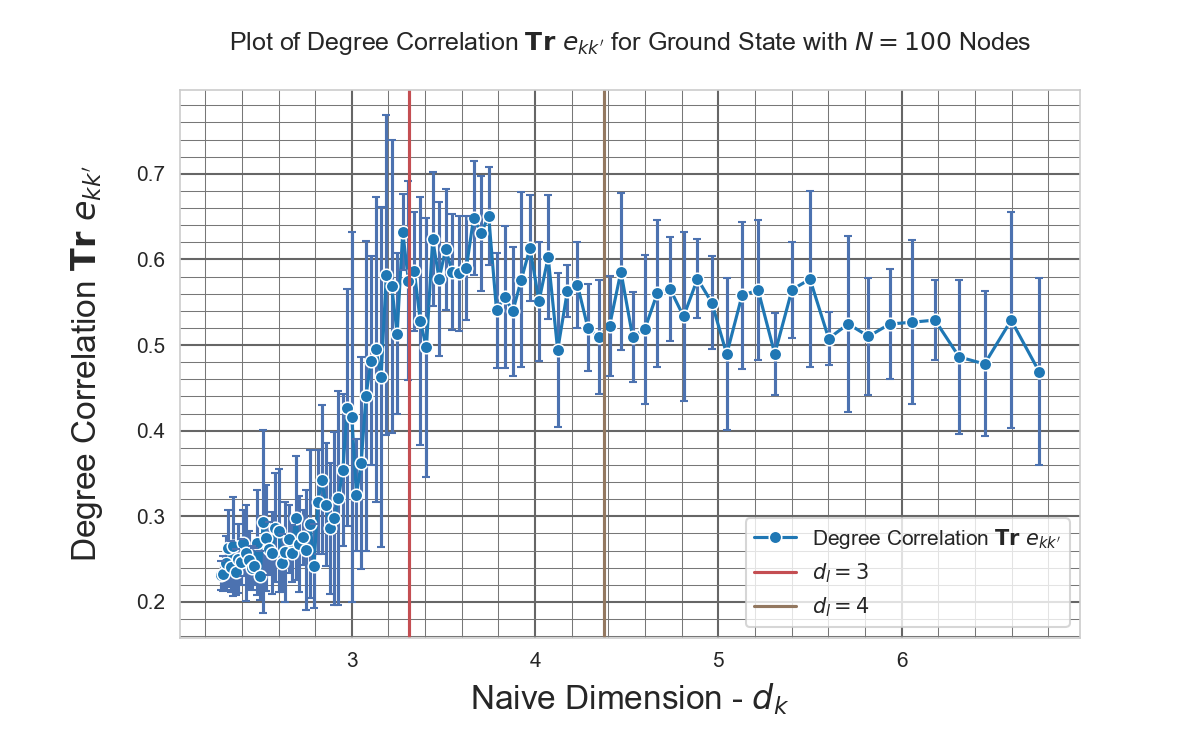}
		\caption{Averaged Trace of Degree Correlation $\Tr e_{kk'}$, $N=100$.}
		\label{fig:TwoPoint100}	
	\end{subfigure}%
	~ 
	\begin{subfigure}[t]{0.45\textwidth}
		\centering
		\includegraphics[scale=0.39]{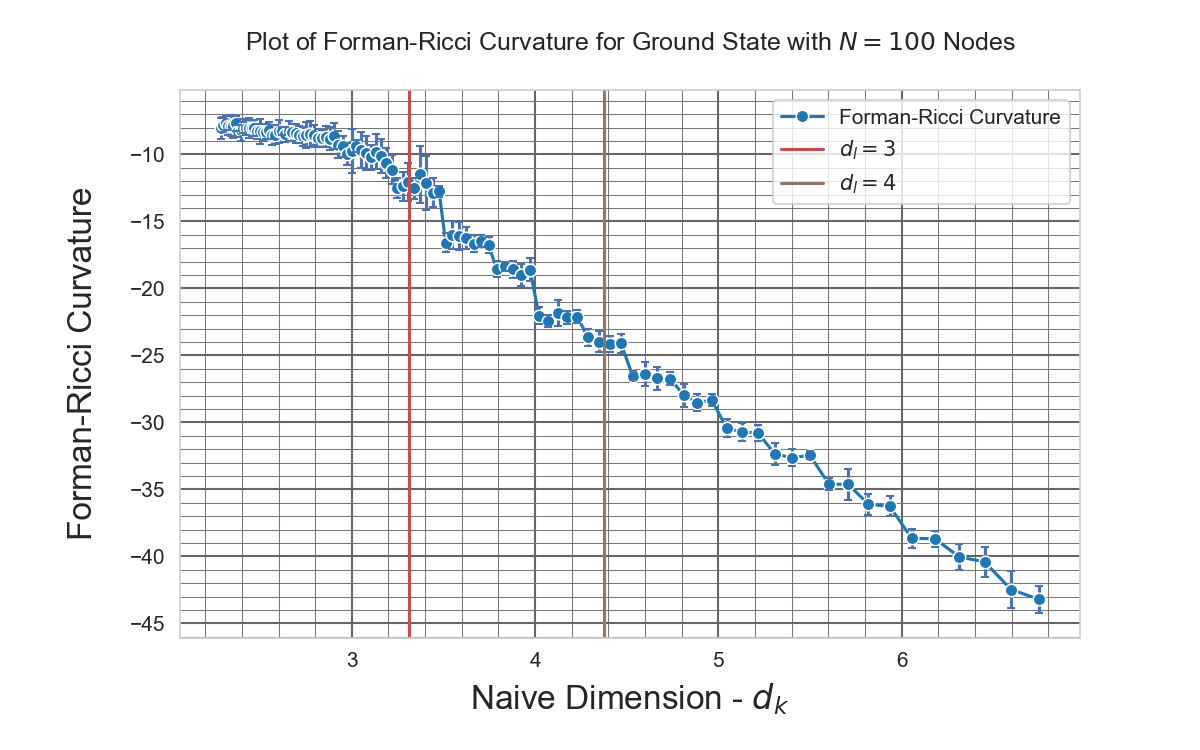}
		\caption{Averaged Forman-Ricci edge curvature, $N=100$.}
		\label{fig:KappaFr100}
	\end{subfigure}
	\caption{Simulation of key ground state metrics for a universe of $N=100$ nodes, using the CIM  Hamiltonian. The coupling constant is varied from $g=0.04$ to $g=0.14$, and the ground state is obtained by minimizing the Hamiltonian in Eq. \eqref{eqn:h_cim_adjacency}. We plot against the naive dimension $d_k$ obtained using equation \Cref{eqn:naive_dimension}.}
	\label{fig:FR-N100CoarseScale}
\end{figure*}

\begin{figure*}[htbp]
	\centering
	\begin{subfigure}[t]{0.45\textwidth}
		\centering
		\includegraphics[scale=0.39]{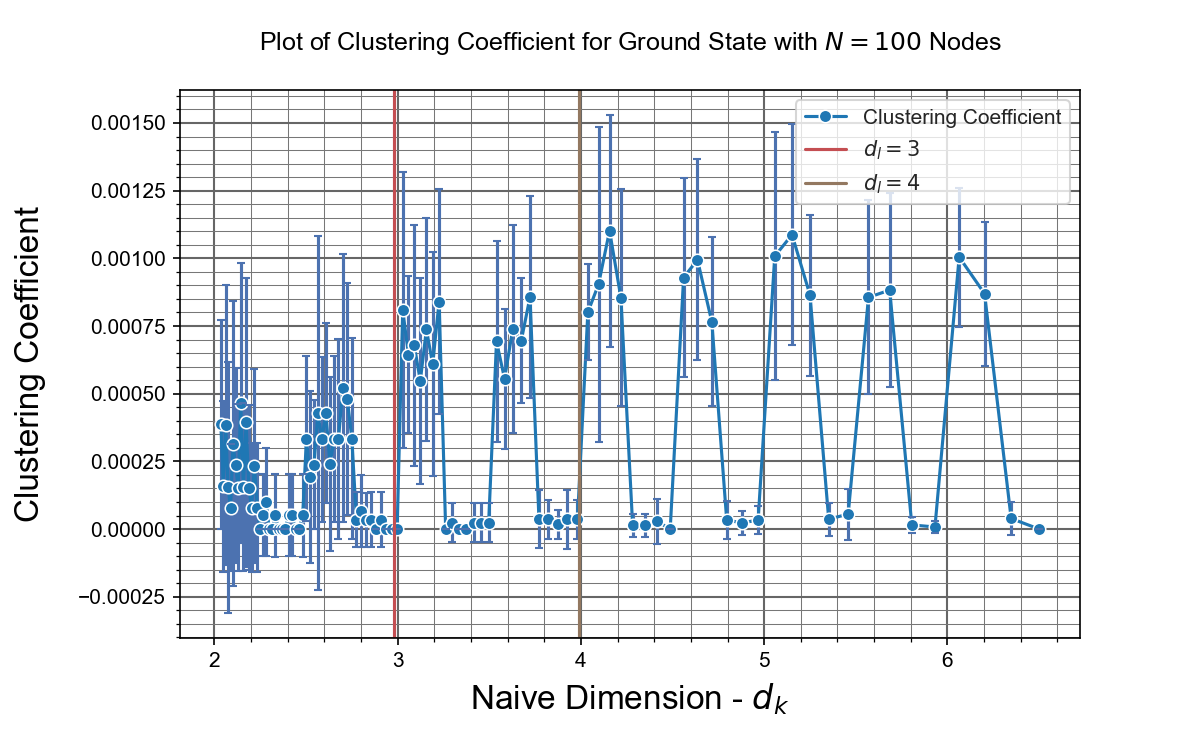}
		\caption{Averaged clustering coefficient, $N=100$.}
		\label{fig:QMD-CC100}
	\end{subfigure}%
	~ 
	\begin{subfigure}[t]{0.45\textwidth}
		\centering
		\includegraphics[scale=0.39]{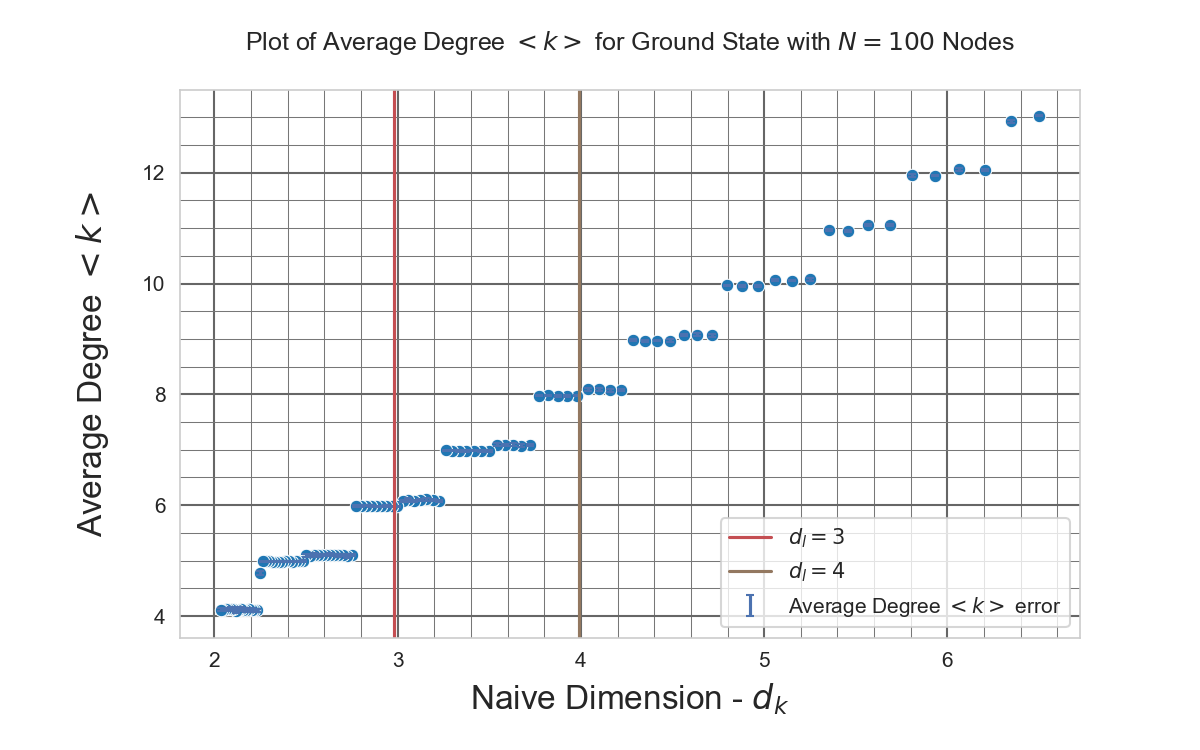}
		\caption{Averaged node degree, $N=100$.}
		\label{fig:QMD-Dgr100}
	\end{subfigure}
	~ 
	\begin{subfigure}[t]{0.45\textwidth}
		\centering
		\includegraphics[scale=0.39]{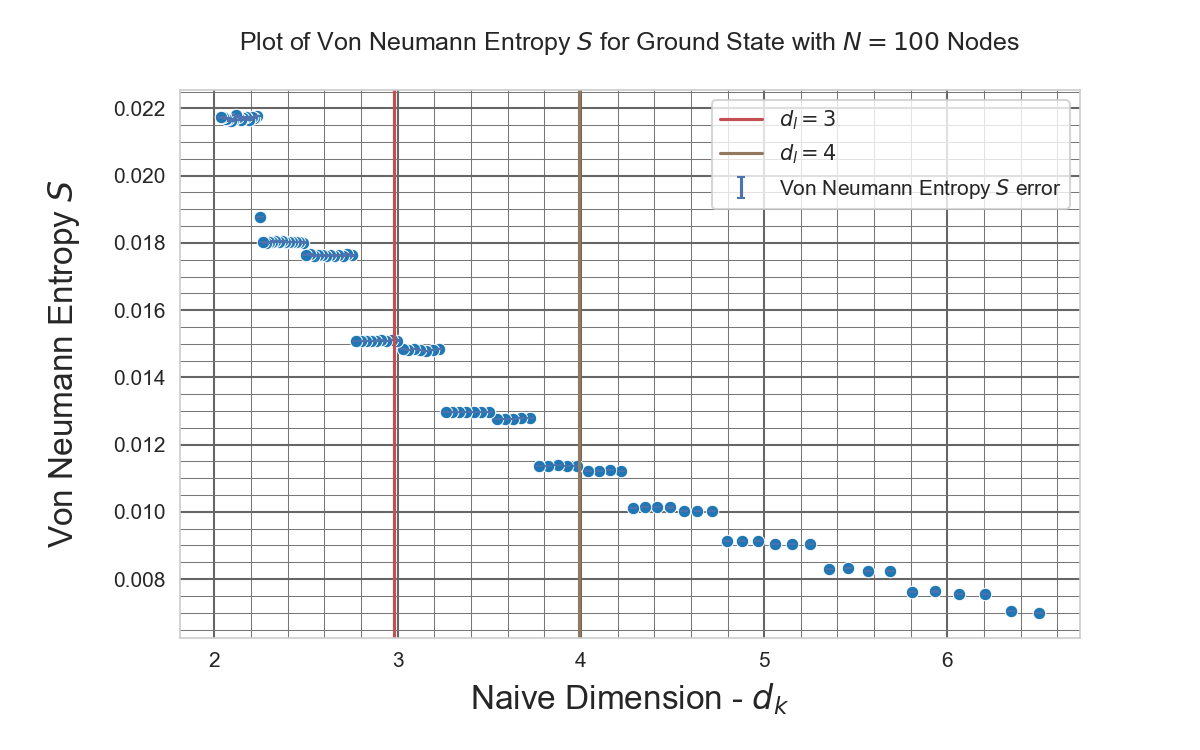}
		\caption{Averaged Von Neumann entropy density, $N=100$.}
		\label{fig:QMD-VNE100}	
	\end{subfigure}%
	~ 
	\begin{subfigure}[t]{0.45\textwidth}
		\centering
		\includegraphics[scale=0.39]{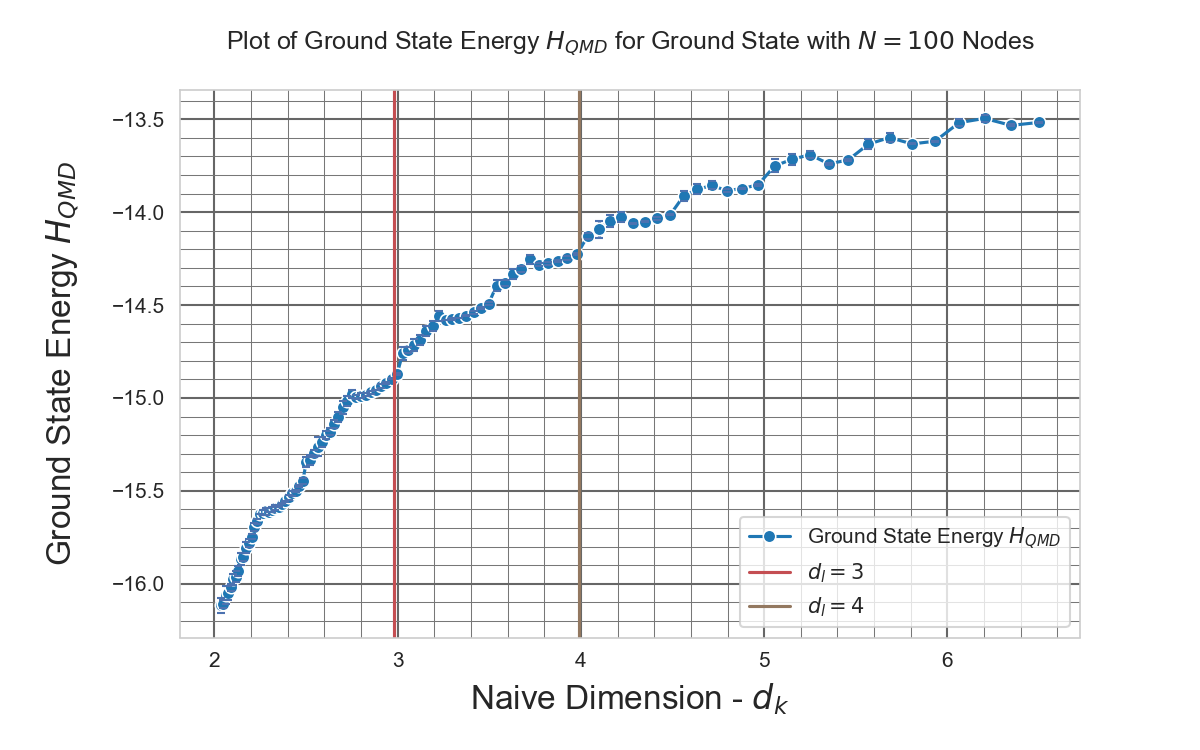}
		\caption{Averaged Graph Hamiltonian, $N=100$.}
		\label{fig:QMD-H100}
	\end{subfigure}
	~
	\begin{subfigure}[t]{0.45\textwidth}
		\centering
		\includegraphics[scale=0.39]{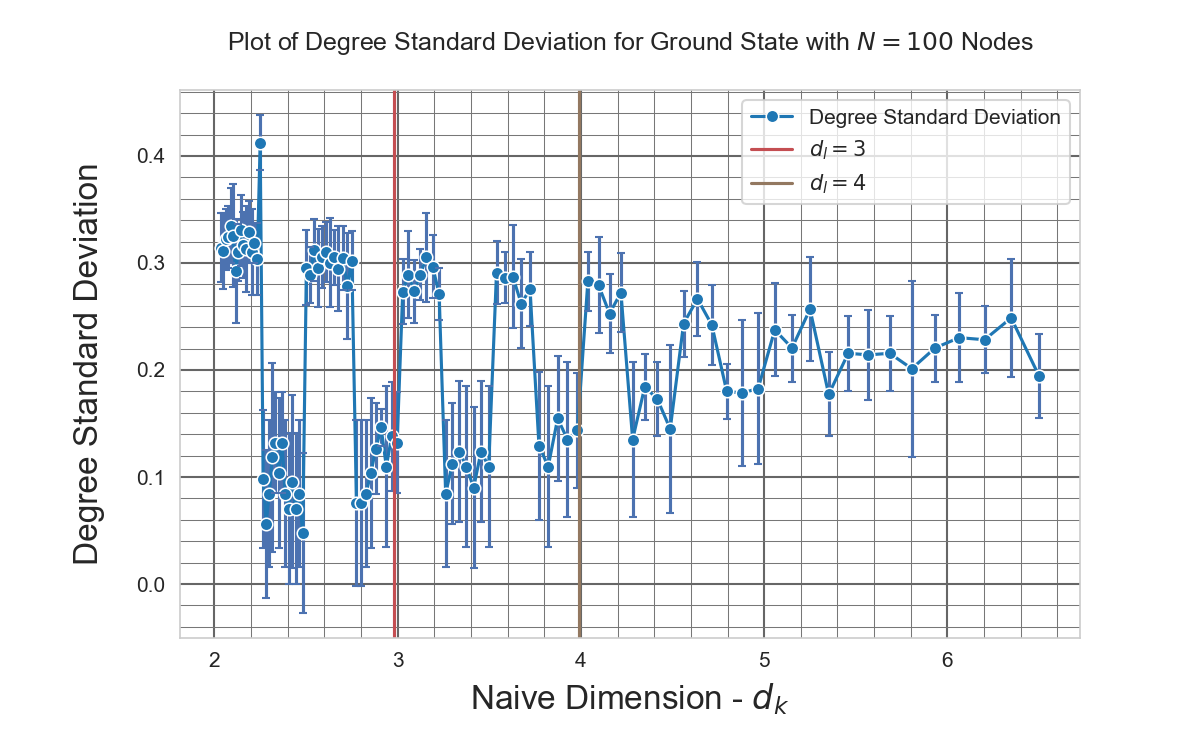}
		\caption{Standard deviation of average node degree, $N=100$.}
		\label{fig:QMD-kStd100}	
	\end{subfigure}%
	~ 
	\begin{subfigure}[t]{0.45\textwidth}
		\centering
		\includegraphics[scale=0.39]{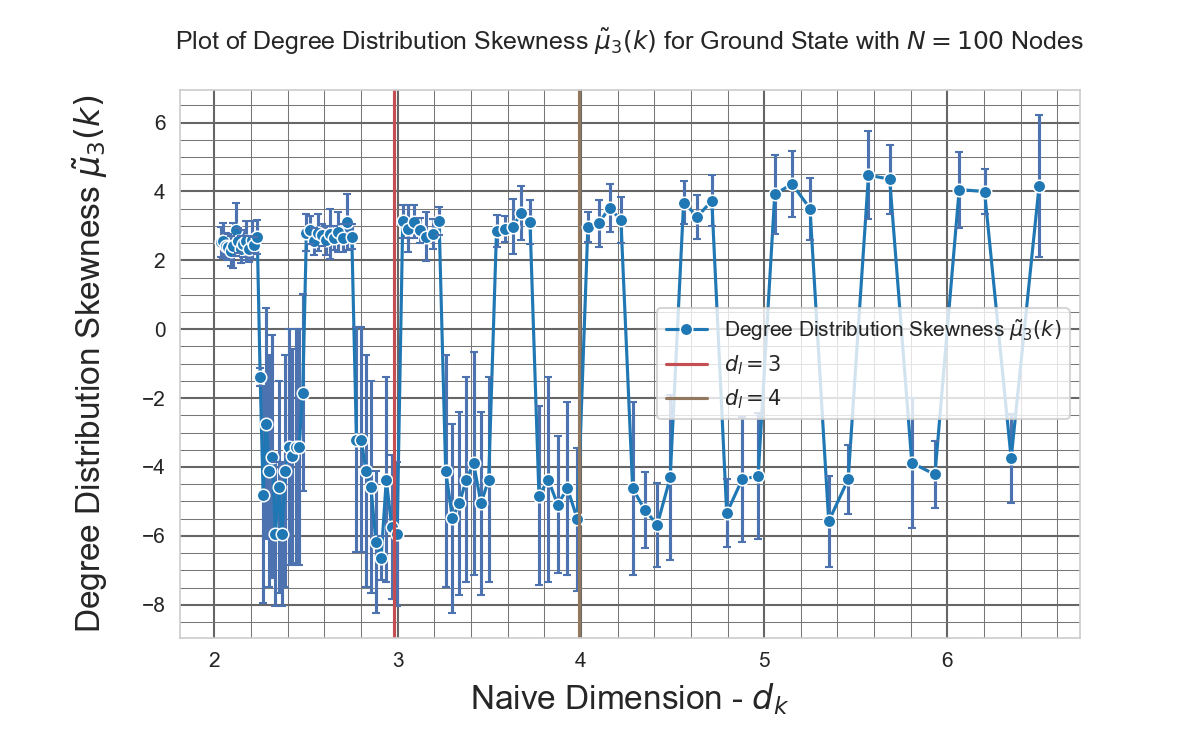}
		\caption{Skewedness of degree distribution, $N=100$.}
		\label{fig:QMD-kSkew100}
	\end{subfigure}
	~ 
	\begin{subfigure}[t]{0.45\textwidth}
		\centering
		\includegraphics[scale=0.39]{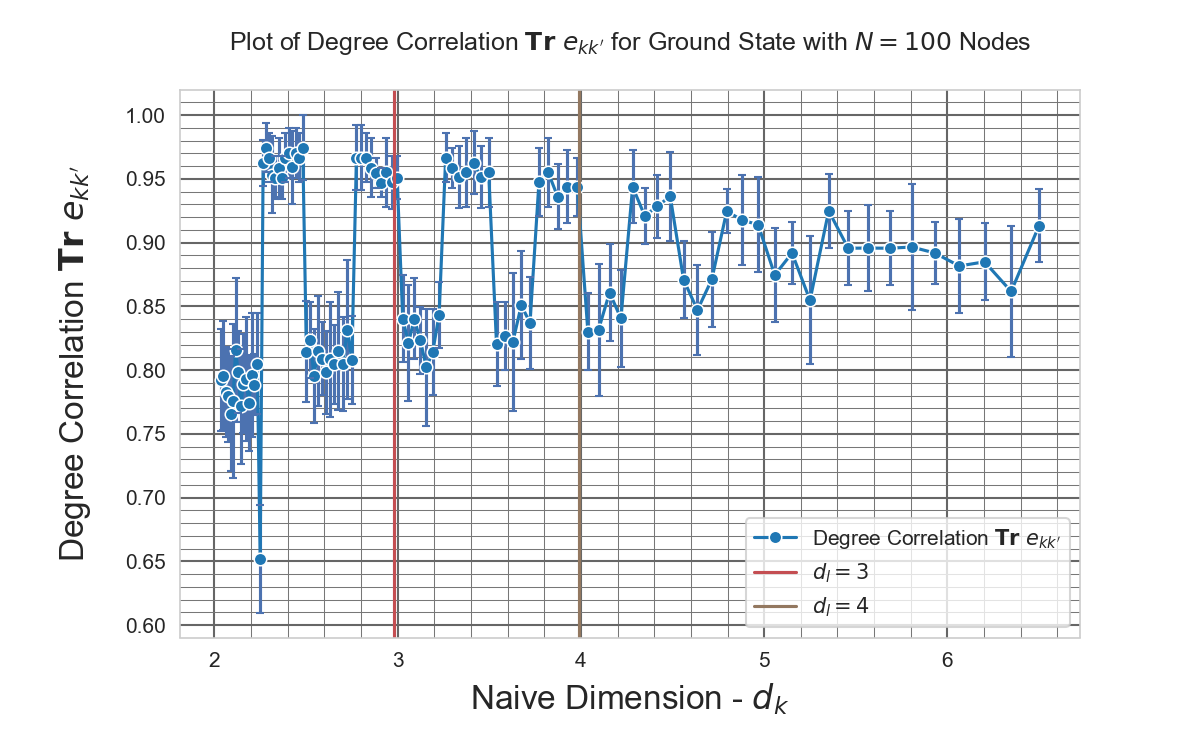}
		\caption{Averaged Trace of Degree Correlation $\Tr e_{kk'}$, $N=100$.}
		\label{fig:QMD-TwoPoint100}	
	\end{subfigure}%
	~ 
	\begin{subfigure}[t]{0.45\textwidth}
		\centering
		\includegraphics[scale=0.39]{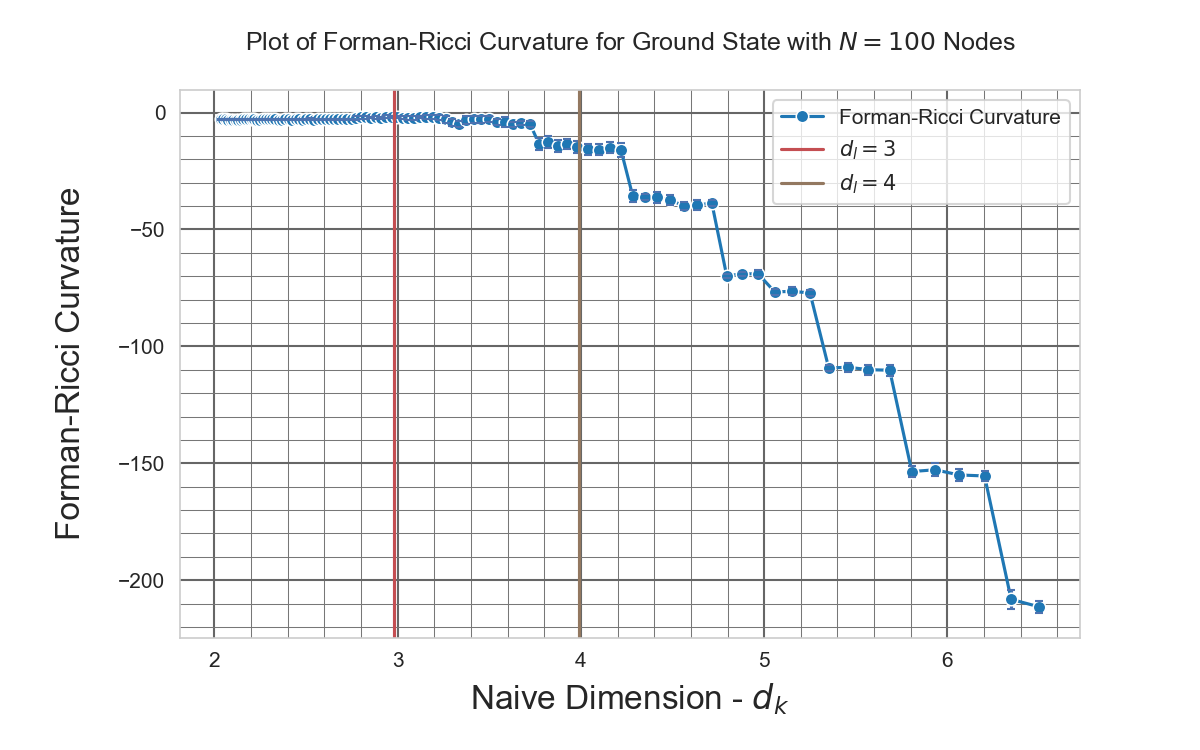}
		\caption{Averaged Forman-Ricci edge curvature, $N=100$.}
		\label{fig:QMD-KappaFr100}
	\end{subfigure}
	\caption{Simulation of key ground state metrics for a universe of $N=100$ nodes, using the QMD Hamiltonian $H_{QMD}$. The coupling constant is varied from $g=0.04$ to $g=0.14$, and the ground state is obtained by minimizing the Hamiltonian in Eq. \eqref{eqn:qmd_hamiltonian}. We plot these against the naive dimension defined in \Cref{eqn:naive_dimension}. The vertical lines represent the value of $d_k$ for which the graph has lattice dimension $d_l=3$ and $d_l=4$.}
	\label{fig:QMD-N100CoarseScale}
\end{figure*}

We present in \Cref{fig:FR-N100CoarseScale} metrics obtained from the ground state graphs of the CIM model defined by \Cref{eqn:h_cim_adjacency}.
We have varied the coupling constant $g$ from $0.04$ to $0.14$, corresponding to a naive dimension in the range of $d_k=2$ to $7$.
We plot the results against naive dimension, but it should be recalled that $d_k$ is inversely proportional to the coupling constant in the graph, and low degree corresponds to larger values of the coupling constant $g$.
In particular it has been suggested \cite{trugenberger2015quantum,trugenberger2017combinatorial,tee2020dynamics} that the coupling constant of Ising models could `run' with the energy or temperature of the system.
It is already visible from \Cref{fig:FR-H100} and \Cref{fig:QMD-H100} that the energy of the system reduces as dimension reduces, or conversely as the coupling constant $g$ increases, corresponding to the  system cooling.
It is possible to interpret the simulations that we have run as representing the state of the spacetime graph at various points in the evolution of the universe from a hotter initial state to a cooler one.
We will address this specific point below.

On each of the graphs we have identified the point where the average node degree corresponds to a lattice dimension of $d_l=3$ and $4$, or alternatively $\expval{k}=6$ and $8$.
We determine these points by identifying the value of $d_k$ where the average degree of the ground state graphs is closest to these values, and plot a vertical line on the figures.

For the results presented in \Cref{fig:FR-N100CoarseScale,fig:QMD-N100CoarseScale,fig:scaled_metrics,fig:ground_state_topo}, the data is presented with the appropriate error bars. 
For the majority of the figures the approach is to use the standard deviation of the results averaged across multiple simulation runs to produce the metric.
Assuming a normal distribution of these results, the bars plotted represent an approximate $90\%$ confidence interval either side of the stated metric.
The exceptions to this approach are the calculations for spectral and Hausdorff dimension, which are computed using a regression fit to the equations \Cref{eqn:spectral_d,eqn:hausdorff}.
For these results the error bars depict a measure dependent upon the $R^2$ coefficient of determination.
This value is identically $1.0$ if the fit is perfect, and our error bars are set to $(1-R^2)$, so a larger bar indicates a less reliable result.
In both of the models considered our coefficient of determination $R^2 > 0.95$, indicating reasonably strong regression fits to the equations defining the dimension measures. 

We collect our observations on the ground state into the following remarks:

\begin{description}
    \item[Locality] In \Cref{fig:CC100} we notice that the clustering coefficient of the obtained graphs is effectively zero for dimensions above $d_l=3$. 
    Below $d_l=3$ the clustering coefficient increases markedly.
    As remarked in earlier work \cite{tee2020dynamics,trugenberger2015quantum}, significant clustering in the graph corresponds to violation of spatial locality.
    This is because triangles introduce the small world property into the spatial graph, allowing points in space to be in contact with potentially distant and non-local ones.
    Locality is a requirement of most theories of physics, at least classically, and it would appear that $d_l=3$ represents a boundary below which the emerged ground state is chaotic and non-local.
    In \Cref{fig:Dgr100} we see that average degree decreases approximately linearly with $d_k$ as we would expect from  \Cref{eqn:naive_dimension} which computes the estimated value of $\expval{k}$ as a function of $g$.
    The fact that the observed value of $\expval{k}$ has a linear relationship with $d_k$ gives us confidence that the assumptions implicit in the calculation of $d_k$ have reasonable validity.
    Just below both $d_l=3$ and $4$ we can detect a sudden drop in $\expval{k}$, which may indicate where our assumptions surrounding the computation of $d_k$ break down.
    We speculate that close to points of high regularity in $\expval{k}$ that this could be caused by a topological phase transition involving a discontinuous change in the number of cycles in the graph.
    Indeed in the plot of cycle density \Cref{fig:FR-polygons} we see that the number of squares incident upon an edge $\expval{\square_{ij}}$ exhibits noticeable discontinuities.
    
    \item[Evolution] Inspecting \Cref{fig:FR-H100} we notice that the value of $H_{CIM}$ decreases for lower values of $d_k$.
    As described earlier, this reduction in energy corresponds to the value of $g$ increasing, and also from the results in \Cref{fig:FR-VNE100} an overall increase in the entropy density of the ground state graphs.
    How do we interpret this?
    We have investigated the ground states of our model across a range of coupling constant values.
    If we accept  that the coupling constant changes as the universe evolves and cools from an original very hot state we arrive at an alternative interpretation of our results as reflecting the emergence of spacetime from an origin event.
    We imagine a universe where energy reduces and entropy increases from such an origin event, in a similar way to the currently accepted standard model of cosmology in which the Universe has evolved from the `Big Bang'.
    As the energy of the universe reduces, from an initial high energy and low entropy state, the dimensionality of it decreases.
    We note that below $d_l=3$ this drop off in energy accelerates, but also from \Cref{fig:CC100} clustering dramatically increases, perhaps indicating that $3$ spatial dimensions is in some way the minimum dimension of our model to exhibit a a highly local geometry.
    This progression from a hot, high-dimensional and low entropy universe to the low dimension, $d_l=3$ state is a process that could be thought of as a 'topological big bang'.
    We assume that the universe has a fixed, very large, number of nodes $N$, consistent with the links having a length in the order of $l_p$. We can then consider how the spatial extent of the universe graph $r_U$ expands as $d_l$ reduces.
    The quantity $r_U$ is the maximum shortest path in the graph, a quantity more usually referred to as the graph's diameter.
    If we assume each link in the spacetime graph represents a `quantum' of distance, this diameter is then physically realized and we can consider it to measure the actual spatial diameter of the ground state graphs.
    From our definition of extrinsic dimension $r_U \propto N^{1/d_l}$, we see that $r_U$ varies with  $d_l$ in a highly non-linear fashion.
    Specifically it implies an initially slow increase in $r_U$ as $d_l$ reduces from the initial very large values of $d_l$, that then accelerates rapidly as dimension reduces towards $3$.
    If we assume that the spatial dimension of the universe does not reduce below $d_l=3$, this acceleration in size will slow down as the universe graph becomes more homogeneous.
    The sudden expansion in the spatial extent of the universe, as $d_l$ reduces, is somewhat analogous to a period of rapid cosmic inflation.
    To recap, interpreting the coupling constant as varying with the energy of the universe, we have an evolution model that starts with a hot, high dimensional and low entropy universe.
    The universe then subsequently undergoes rapid spatial expansion that slows down as it cools, entropy increases and spatial dimension reduces to $d_l=3$.
    
    \item[Regularity] In \Cref{fig:kStd100} we plot the standard deviation of the average node degree.
    It is notable that in general this reduces with $d_k$ towards a minimum value at $d_l=3$, at which point it increases rapidly.
    This indicates that the ground state graphs are steadily getting more regular and uniform up until that point.
    This is consistent with the results in \Cref{fig:TwoPoint100} for the two-point correlation function $\Tr e_{kk'}$.
    The value of $\Tr e_{kk'}$ measures the probability that a randomly chosen link connects nodes of equal degree.
    The value of $\Tr e_{kk'}$ peaks around $d_l=3$, indicating that at this dimension the ground state graphs become maximally uniform.
    The value, however, is not identically equal to $1.0$, which one would obtain for a perfectly regular lattice.
    In \Cref{fig:kSkew100} we plot the third moment of the degree distribution, $\tilde{\mu}_3(k)$ the `skewness'.
    In a small interval surrounding $d_l=3$ the distribution switches from a positive to a negative skew.
    When the distribution has $\tilde{\mu}_3(k) < 0$ the distribution is described as `left-tailed' and indicates the population of nodes with degree below the mean value is larger than the population above the mean.
    It is generally accepted that values of $\tilde{\mu}_3(k)$ in the range $[-0.5,0.5]$ are consistent with a symmetrical distribution and the significance is only important for $\tilde{\mu}_3(k) >1$ and $\tilde{\mu}_3(k) < -1$.
    Using this interpretation, the switch from positive to negative skew in our ground states would appear significant.
    This could potentially indicate the emergence of a boundary to the graph that would naturally entail a collection of nodes of lower degree than the bulk, particularly if the number of nodes with lower degree is significant and peaked around a second local maximum.
    If this is not the case the skew could simply occur because of the presence of low connectivity defects.
    For values of $d_k$ corresponding to positive skew, it could be as a results of the presence of defects involving regions of higher connectivity.
    A more detailed investigation of the structure of the graph is required to definitively identify the cause of this variability in skew, but this is the subject of future work.
    
    \item[Curvature] We plot in \Cref{fig:KappaFr100} the average FR curvature of an edge in the ground state graphs.
    In general the total curvature of the ground state graphs are negative, but as they  lower in dimension the average edge curvature increases towards zero.
    The computation in \Cref{sec:discreteGR} relied upon a choice of `gauge' for the canonical Hamiltonian, which amounted to setting the lapse and shift vectors for our foliation to $N=1$, and $N_i=0$. 
    As these parameters are not dynamical variables in the canonical GR formulation, this freedom comes with two constraints on the system, usually applied to the system's initial conditions.
    These are referred to as the Hamiltonian and momentum constraints.
    They are enforced through the definition of two quantities,
    \begin{equation}
    \begin{split}
        \mathcal{C}&=~^3R+K^2+K^{ij}K_{ij}, \\
        \mathcal{C}_i&=(K_i^j-K\delta_i^j)_|j \text{,}
    \end{split}
    \end{equation}
    which must both be zero.
    We recall that we identified $\kappa^f$ with $~^3R$ to obtain our Hamiltonian. 
    Following that analogy, the first of these implies that as $\kappa^f < 0$, the terms involving the extrinsic curvature must satisfy $K^2-K^{ij}K_{ij} > 0$, and both will tend to zero as the dimension reduces and the graph lowers in energy.
    We argue that in the continuum limit assuming $\kappa^f$ corresponds to $~^3R$, and the relationship between the extrinsic and intrinsic curvatures implies that the spacetime Ricci curvature of the hypersurface is, to a first approximation, zero.
    Indeed for smooth manifolds the Gauss-Codazzi equations \cite{poisson2004relativist} relate the Ricci scalar for hypersurfaces to $\mathcal{C}$ and terms in the second order covariant derivative of the normal vectors to the hypersurfaces  $\Sigma_t$ we consider.
    So if we stretch our analogy associating the FR curvature to the intrinsic curvature of spacelike hypersurfaces, our results imply that for low dimensions we have an approximately flat spacetime geometry.
    This is consistent with current experimental observations of the curvature of the Universe \cite{xia2017revisiting,handley2021curvature}, with the current consensus being that the spatial curvature is either zero or weakly positive.
\end{description}

We compare these results with those obtained by minimizing the QMD Hamiltonian defined in \Cref{eqn:qmd_hamiltonian}.
In general the results are mostly similar to those obtained using the CIM model.
Specifically we note:

\begin{description}
    \item[Clustering, Connectivity and Regularity] In \Cref{fig:QMD-CC100} we plot the clustering coefficient, noting that the scale is considerably compressed compared to the results for the CIM model in \Cref{fig:CC100}. 
    In general the ground state is highly local, with a general trend towards lower  clustering as $d_k$ reduces, but with periodic elevated clustering between regions of none.
    The clustering coefficient however, does not exhibit a sudden increase for values of $d_k<3$, as seen in the $H_{CIM}$ model.
    We also note that the vertical lines placed at $d_l =3$, and $4$ correspond to regimes of effectively zero clustering.
    Additionally in \Cref{fig:QMD-Dgr100} we note a much more coarse, but highly linear relationship between $\expval{k}$ and $d_k$.
    The correspondence between $d_l$ and $d_k$ is almost exact, reflecting the fact that the computation of $d_k$ for the QMD model does not involve the approximations necessary to analytically minimize the CIM Hamiltonian.
    The distribution of node degree demonstrates some important differences to CIM.
    In \Cref{fig:QMD-kappaStd} the standard deviation of the degree distribution is considerably tighter in QMD than CIM, and again the lattice dimensions of $3$ and $4$ correspond to near zero values of standard deviation.
    This is repeated in \Cref{fig:QMD-TwoPoint100} where the values of $\Tr (e_{kk'})$ approach unity at $d_l=3$.
    The third moment plotted in \Cref{fig:QMD-kSkew100} is also highly negative, indicating that any deviation involves a small number of nodes with degree less than $\expval{k}$.
    This could be interpreted as the presence of a boundary, but we believe that this interpretation is much weaker for QMD than for CIM, given how small the variation in $\expval{k}$ is in QMD.
    Indeed this could simply be due to experimental error, if we take into account the average degree variation.
    Comparing the degree variation in \Cref{fig:kStd100} and \Cref{fig:QMD-kStd100}, we see that the QMD model is much more uniform, and the presence of a boundary would require a degree distribution highly peaked around two values, $\expval{k}$ and $(\expval{k}-1)$.
    Further investigation is required for both models.
    
    \item[Evolution and Curvature] The results for entropy \Cref{fig:QMD-VNE100} and energy \Cref{fig:QMD-H100} are both comparable with the CIM model results.
    Perhaps one could argue that the relationship is less smooth than in the CIM model, but this is a function of the stepped nature of the dependence of $\expval{k}$ on $d_k$.
    We have already remarked that the CIM Hamiltonian is very similar to QMD, with the additional terms indicating that QMD is an approximation of CIM.
    The coarseness of the QMD results, with the presence of step discontinuities in many of the metrics, could be a consequence of this.
    This discontinuous variation is in even more evidence with the curvature distribution in  \Cref{fig:QMD-KappaFr100}.
    The curvature still tends upwards to zero with decreasing $d_k$ and we can make the same observations as were made for CIM, that the Hamiltonian constraints imply a Ricci flat ground state.
    Indeed at $d_l=3$ the curvature of the ground state graph is near zero, indicating that the QMD ground states are spatially flat.
\end{description}

\subsection{Small scale effects}

\begin{figure*}[htbp]
	\centering
	\begin{subfigure}[t]{0.45\textwidth}
		\centering
		\includegraphics[scale=0.39]{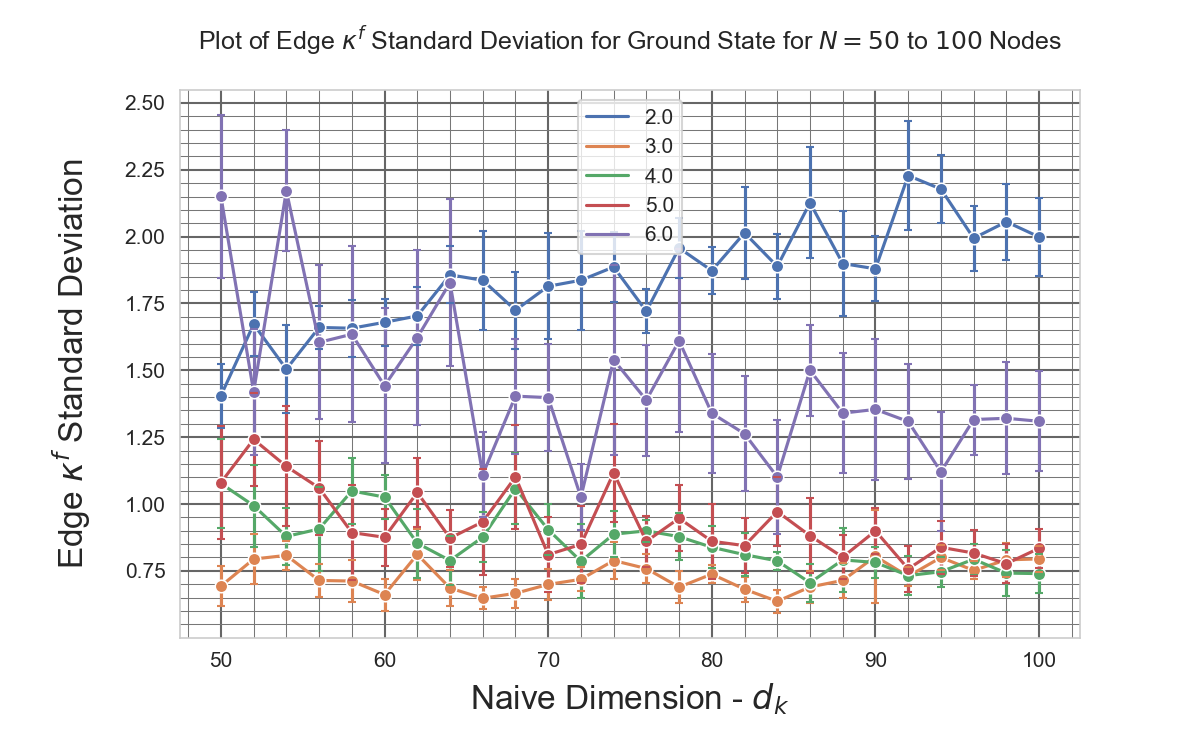}
		\caption{Standard deviation of Forman-Ricci Curvature $H_{CIM}$.}
		\label{fig:FR-ScaledKappa}
	\end{subfigure}%
	~ 
	\begin{subfigure}[t]{0.45\textwidth}
		\centering
		\includegraphics[scale=0.39]{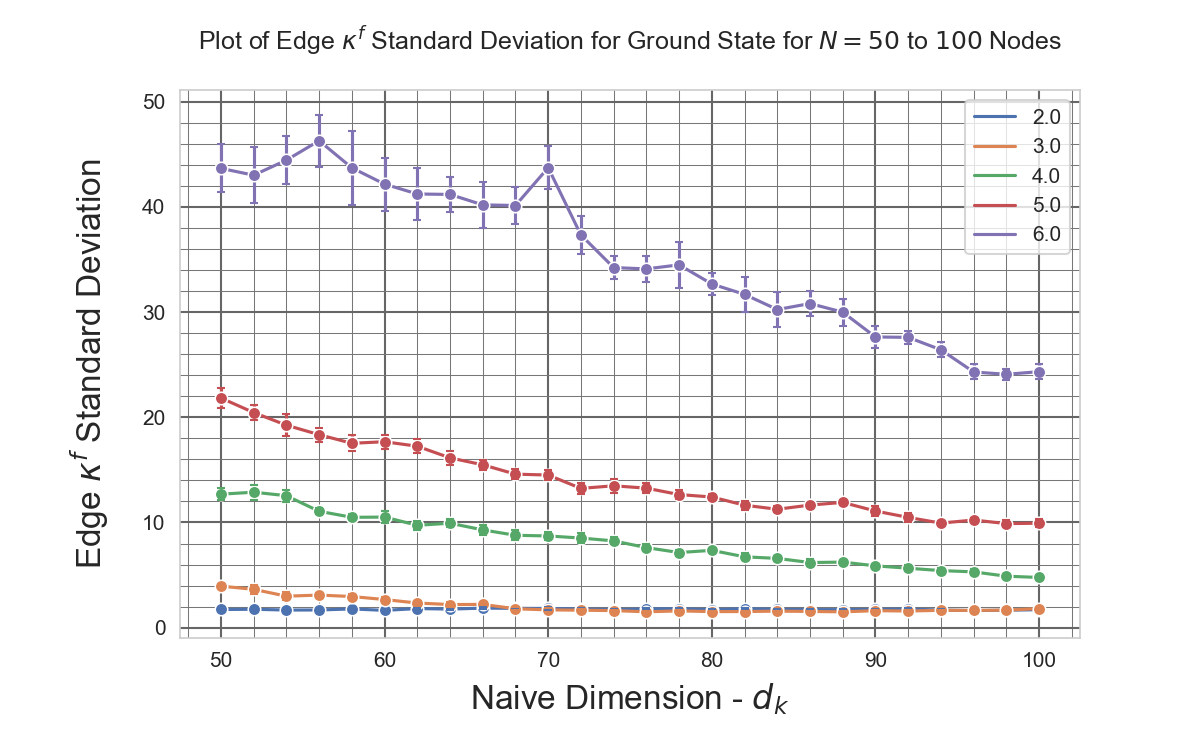}
		\caption{Standard deviation of Forman-Ricci Curvature $H_{QMD}$.}
		\label{fig:QMD-ScaledKappa}
	\end{subfigure}
	~ 
	\begin{subfigure}[t]{0.45\textwidth}
		\centering
		\includegraphics[scale=0.39]{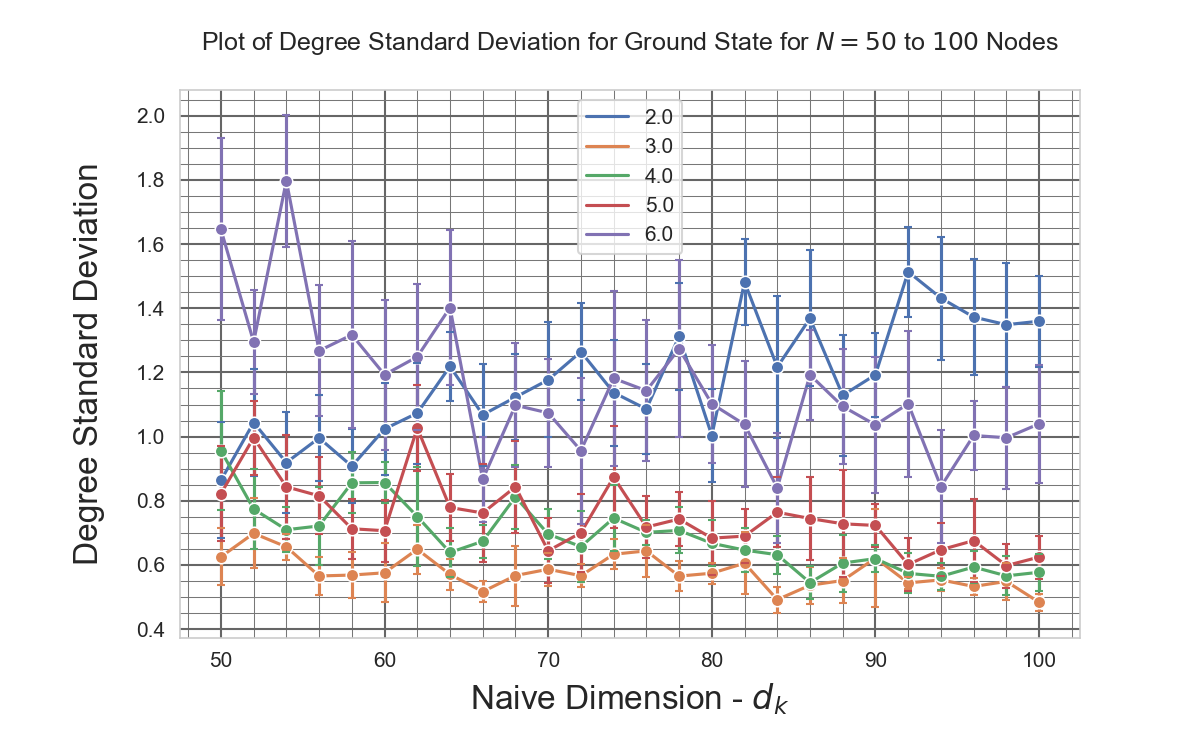}
		\caption{Standard deviation of node degree $H_{CIM}$.}
		\label{fig:FR-ScaledkStd}	
	\end{subfigure}%
	~ 
	\begin{subfigure}[t]{0.45\textwidth}
		\centering
		\includegraphics[scale=0.39]{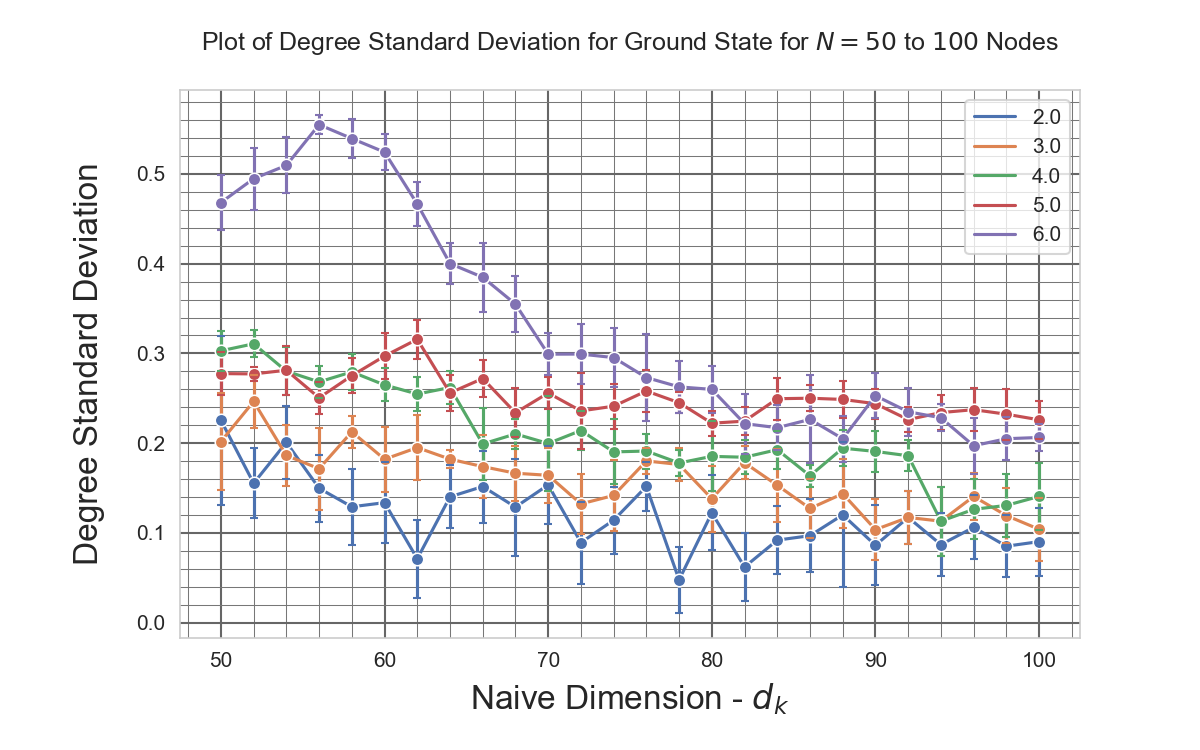}
		\caption{Standard deviation of node degree $H_{QMD}$.}
		\label{fig:QMD-ScaledkStd}
	\end{subfigure}
	~
	\begin{subfigure}[t]{0.45\textwidth}
		\centering
		\includegraphics[scale=0.39]{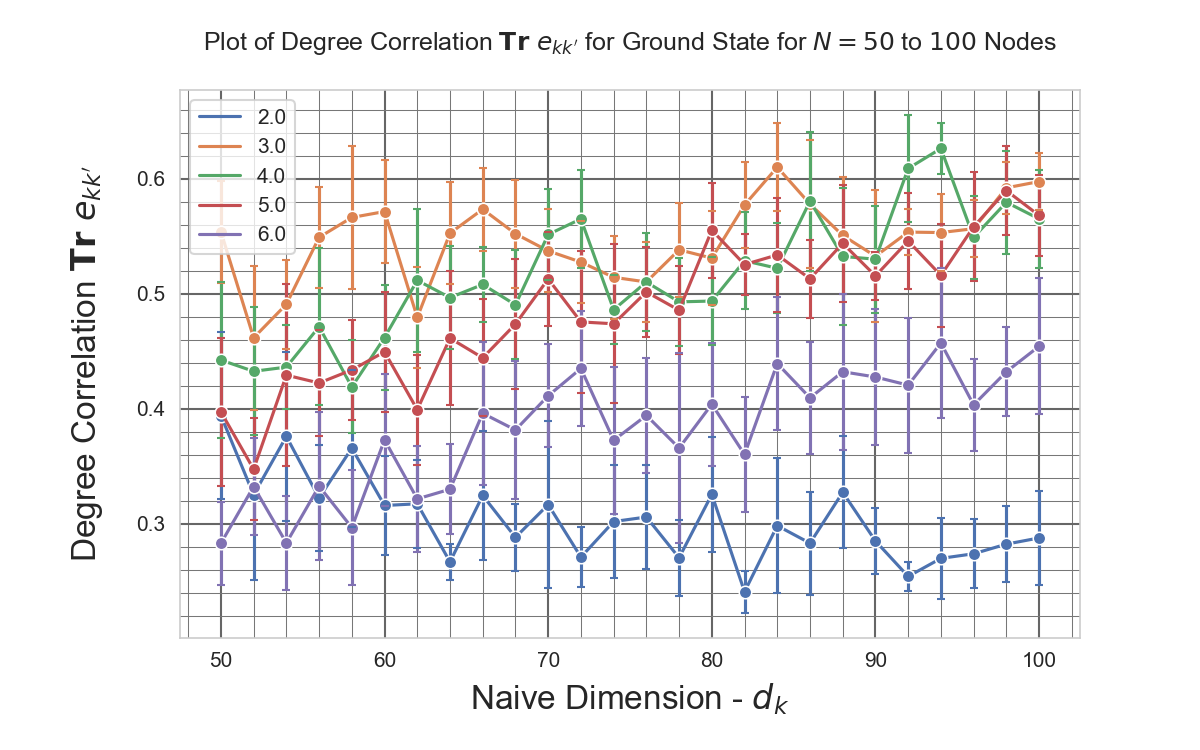}
		\caption{Two-point correlation function for  $H_{CIM}$.}
		\label{fig:FR-ScaledTwoPoint}	
	\end{subfigure}%
	~ 
	\begin{subfigure}[t]{0.45\textwidth}
		\centering
		\includegraphics[scale=0.39]{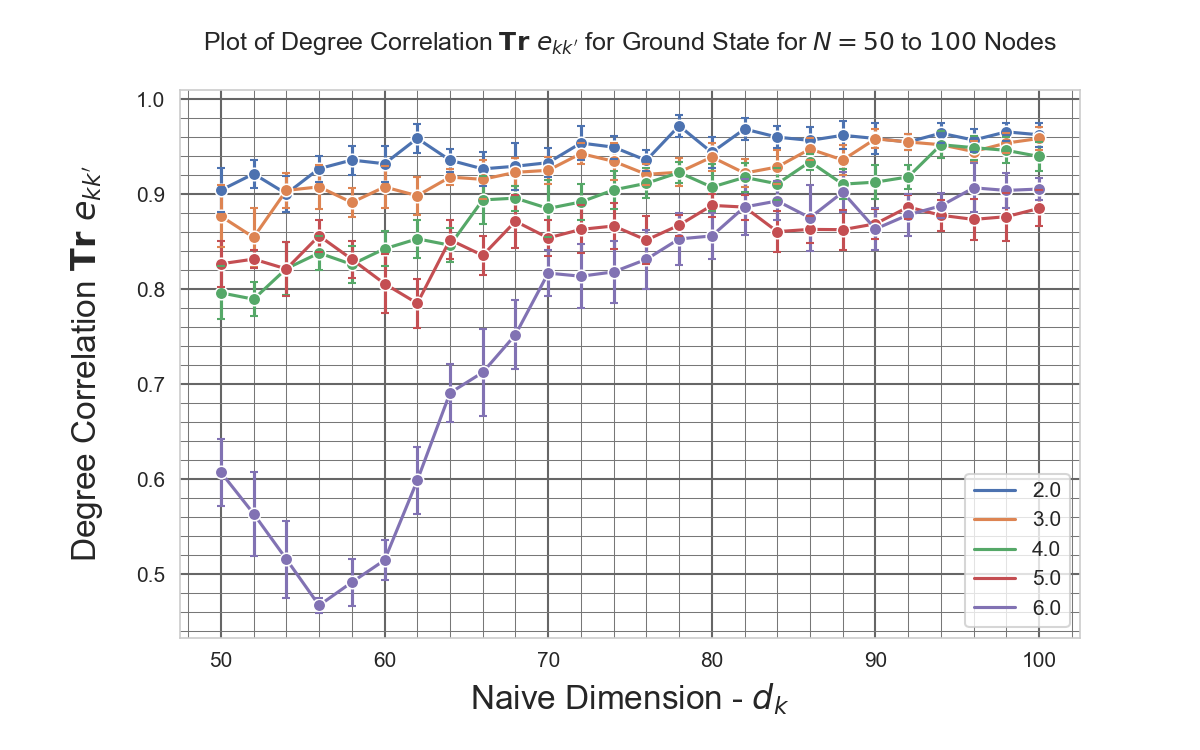}
		\caption{Two-point correlation function for  $H_{QMD}$.}
		\label{fig:QMD-ScaledTwoPoint}
	\end{subfigure}
	\caption{Using the Hamiltonians for the QMD and CIM models we present a selection of  metrics, at  fixed values of dimension, for graphs of sizes ranging from $N=50$ to $N=100$. The metrics are averaged over a collection of $10$ runs for each size of graph and we plot the value of the metrics for a selection of naive dimensions $d_k=2 \text{, to~}6$. The vertical lines represent the value of $d_k$ for which the graph has lattice dimension $d_l=3$ and $d_l=4$.}
	\label{fig:scaled_metrics}
\end{figure*}

Computational limitations prevent us from considering very large graphs in our numerical simulations.
We have investigated the scaling dependence of some of our metrics on graph size, and in \Cref{fig:scaled_metrics} we present a subset of the results for selected metrics for both models, varying graph size from $N=50$ to $100$.
The subset of metrics is selected to highlight those that are particularly sensitive to boundary features and scale effects.

In \Cref{fig:FR-ScaledKappa} and \Cref{fig:QMD-ScaledKappa} we plot the standard deviation of the edge FR curvature. 
This is a key measure of regularity in the ground state and in both cases it reduces as the graphs increase in size.
The only exception to this is in the case of the CIM models for $d_k<3$ where the graph becomes more irregular.
We will comment more in the following section but for low values of $d_k$ the ground state graph becomes chaotic, unstable and disordered.
This is characterized by a pronounced increase in clustering and a drop in regularity of the structure of the graph.

In \Cref{fig:FR-ScaledkStd} and \Cref{fig:QMD-ScaledkStd} we plot the standard deviation of node degree.
Both results are similar, and show increasingly regular graphs as their size increases.
Once more in the case of CIM, for $d_k < 3$ the graphs become more irregular, consistent with there being a transition to a much more disordered ground state below $d_k=3$.
The drop in the standard deviation of node degree with graph size could occur if there is a boundary feature in the topology of the graph, which would of course consume fewer nodes as a proportion of the total, as the graph scales.

Turning to degree correlation in \Cref{fig:FR-ScaledTwoPoint} and \Cref{fig:QMD-ScaledTwoPoint} we see a similar trend with nodes becoming more assortative and exhibiting higher degree correlation, with the exception of CIM for the graphs with $d_k<3$.

We conclude that the graph's regularity increases with size, which would support the expectation that conclusions we have drawn regarding the structure of our ground state graphs are likely to have validity for much larger graphs. 

\subsection{Topology, regularity and dimension}

\begin{figure*}[htbp]
	\centering
	\begin{subfigure}[t]{0.45\textwidth}
		\centering
		\includegraphics[scale=0.38]{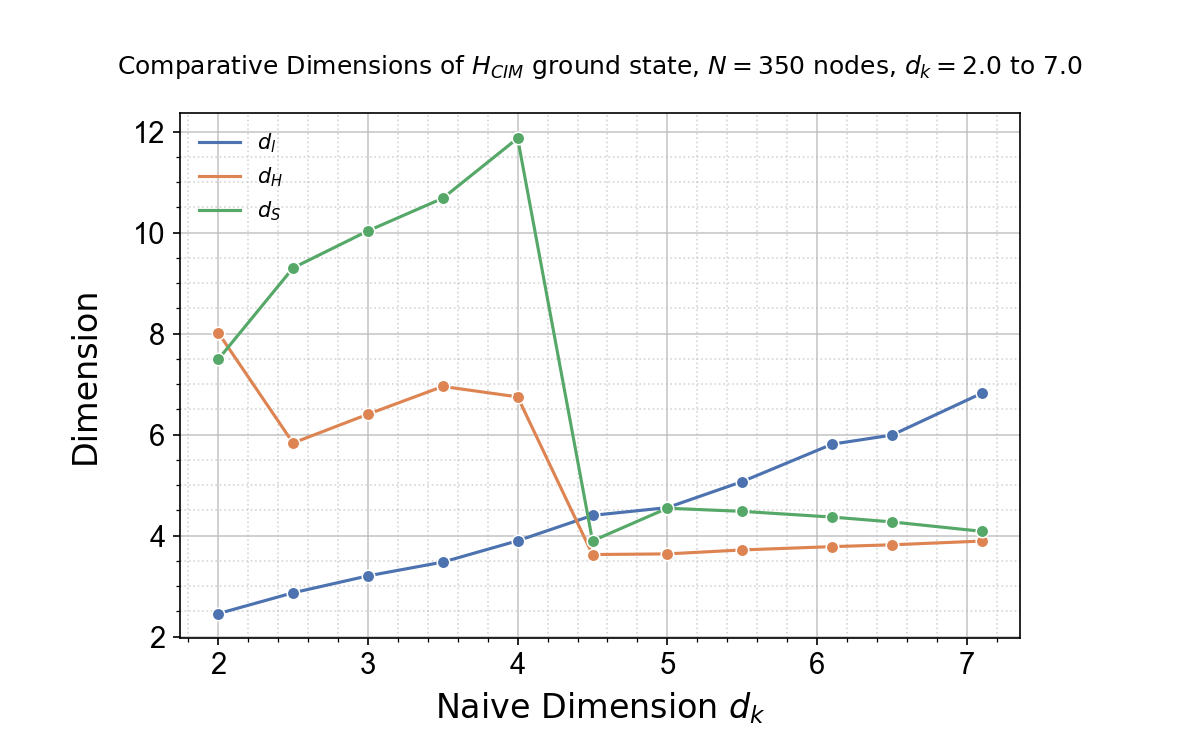}
		\caption{Intrinsic, extrinsic and naive dimension $N=350$, for $H_{CIM}$. Error bars are plotted but not visible as all values are below $0.05$.}
		\label{fig:FR-Dims}
	\end{subfigure}%
	~ 
	\begin{subfigure}[t]{0.45\textwidth}
		\centering
		\includegraphics[scale=0.38]{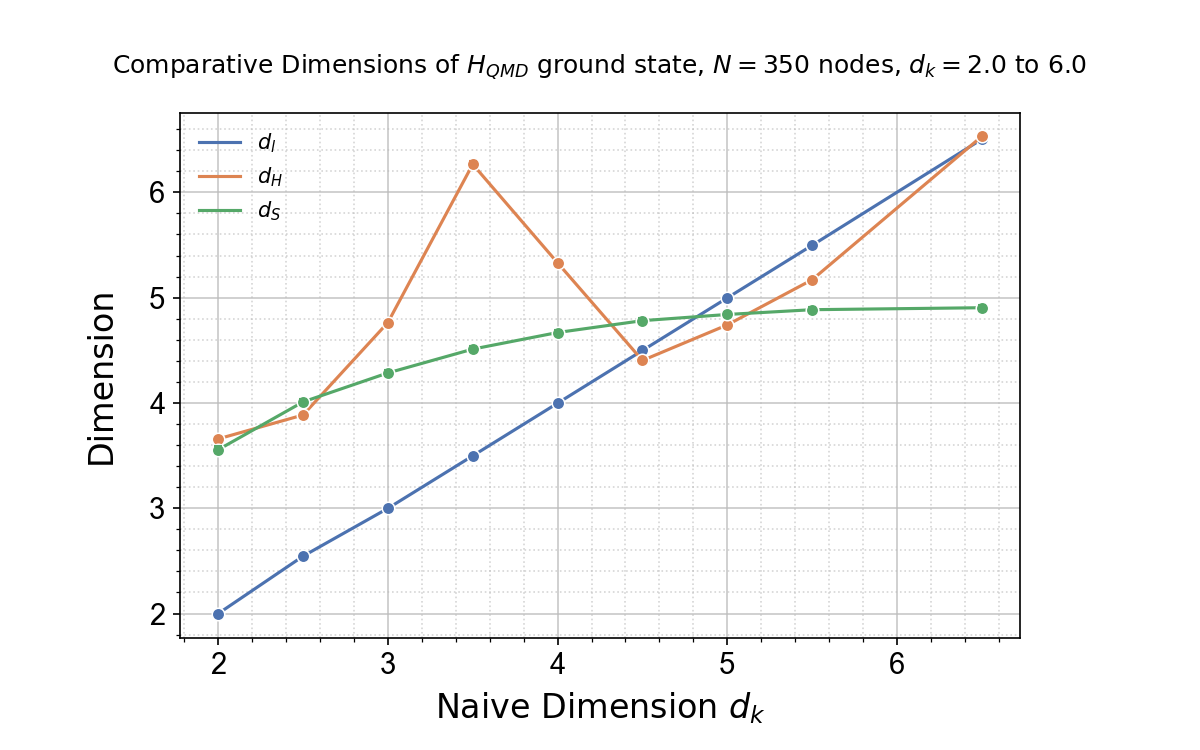}
		\caption{Intrinsic, extrinsic and naive dimension $N=350$, for $H_{QMD}$. Error bars are plotted but not visible as all values are below $0.05$.}
		\label{fig:QMD-Dims}
	\end{subfigure}
	~ 
	\begin{subfigure}[t]{0.45\textwidth}
		\centering
		\includegraphics[scale=0.38]{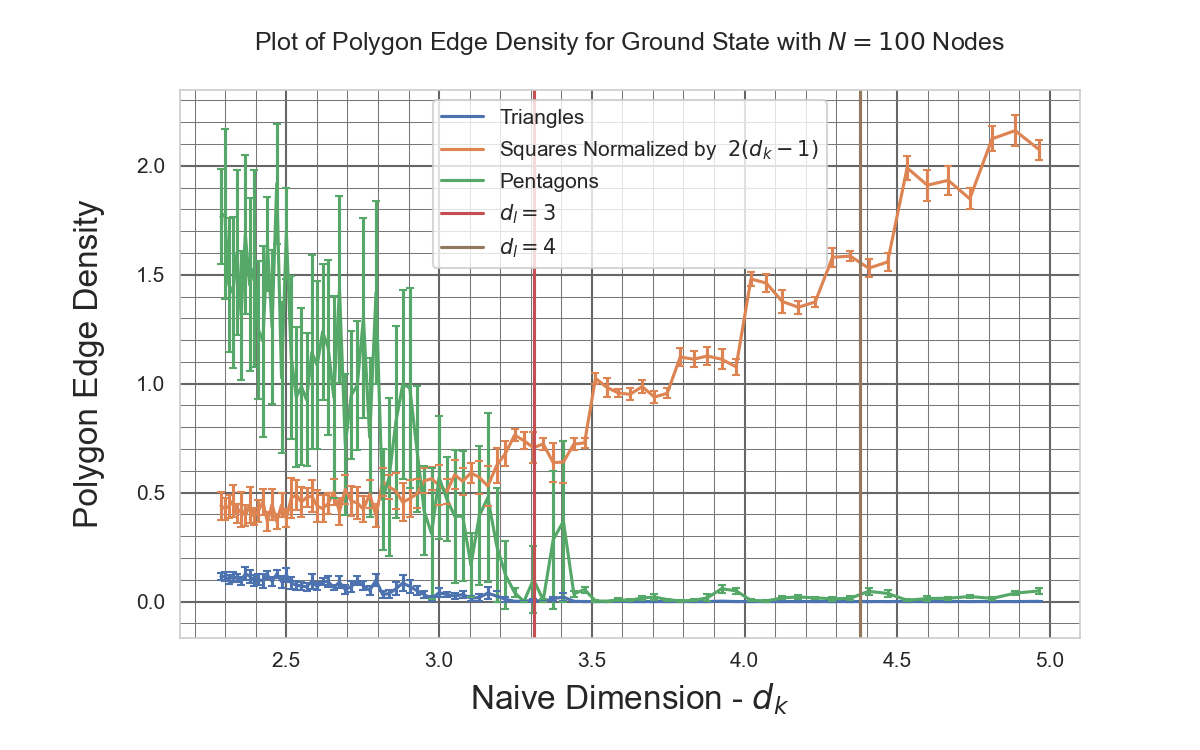}
		\caption{Cycle count versus naive dimension, for $H_{CIM}$ and $N=350$.}
		\label{fig:FR-polygons}	
	\end{subfigure}%
	~ 
	\begin{subfigure}[t]{0.45\textwidth}
		\centering
		\includegraphics[scale=0.38]{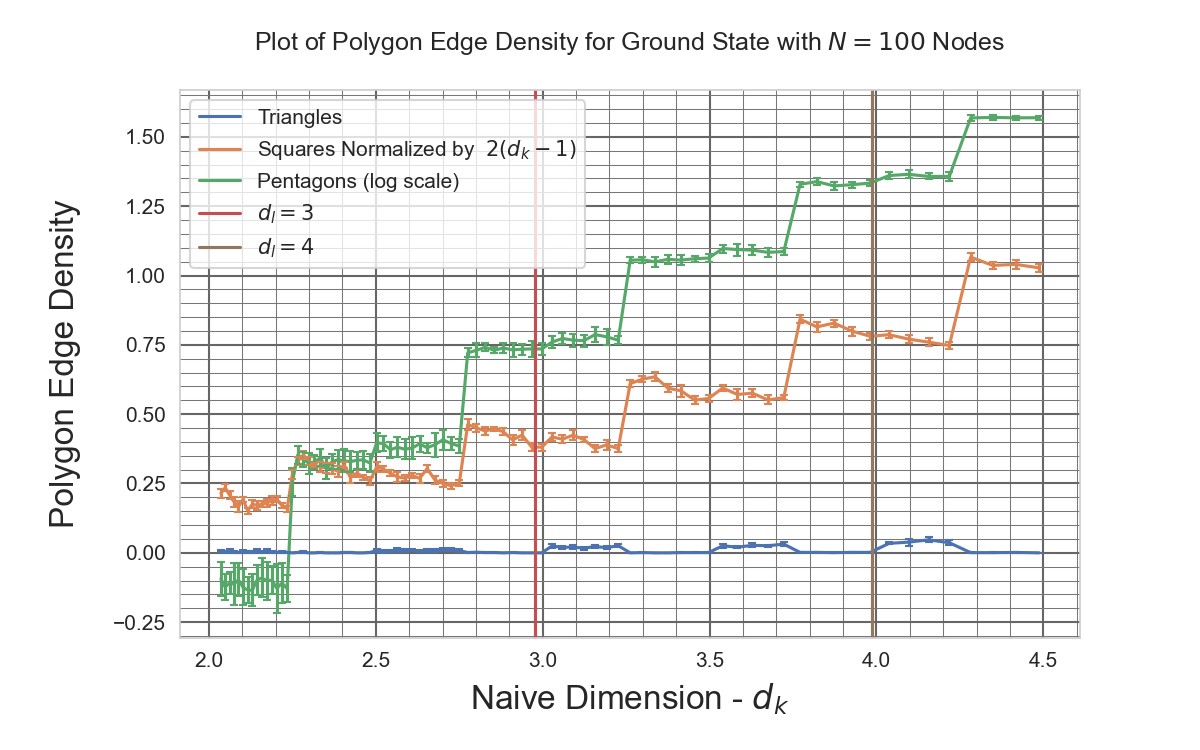}
		\caption{Cycle count versus naive dimension, for $H_{QMD}$ and $N=350$.}
		\label{fig:QMD-polygons}
	\end{subfigure}
	~
	\begin{subfigure}[t]{0.45\textwidth}
		\centering
		\includegraphics[scale=0.38]{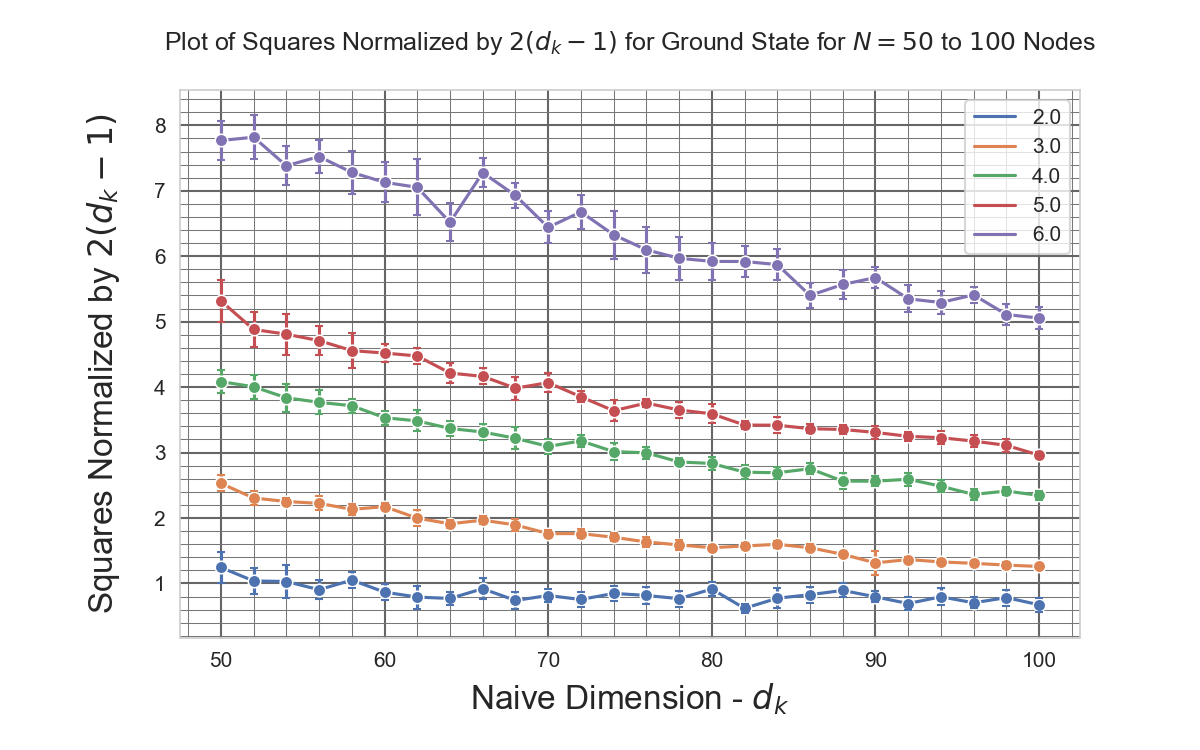}
		\caption{Normalized square cycle density versus naive dimension, for $H_{CIM}$ and $N=350$.}
		\label{fig:FR-sqDens}	
	\end{subfigure}%
	~ 
	\begin{subfigure}[t]{0.45\textwidth}
		\centering
		\includegraphics[scale=0.38]{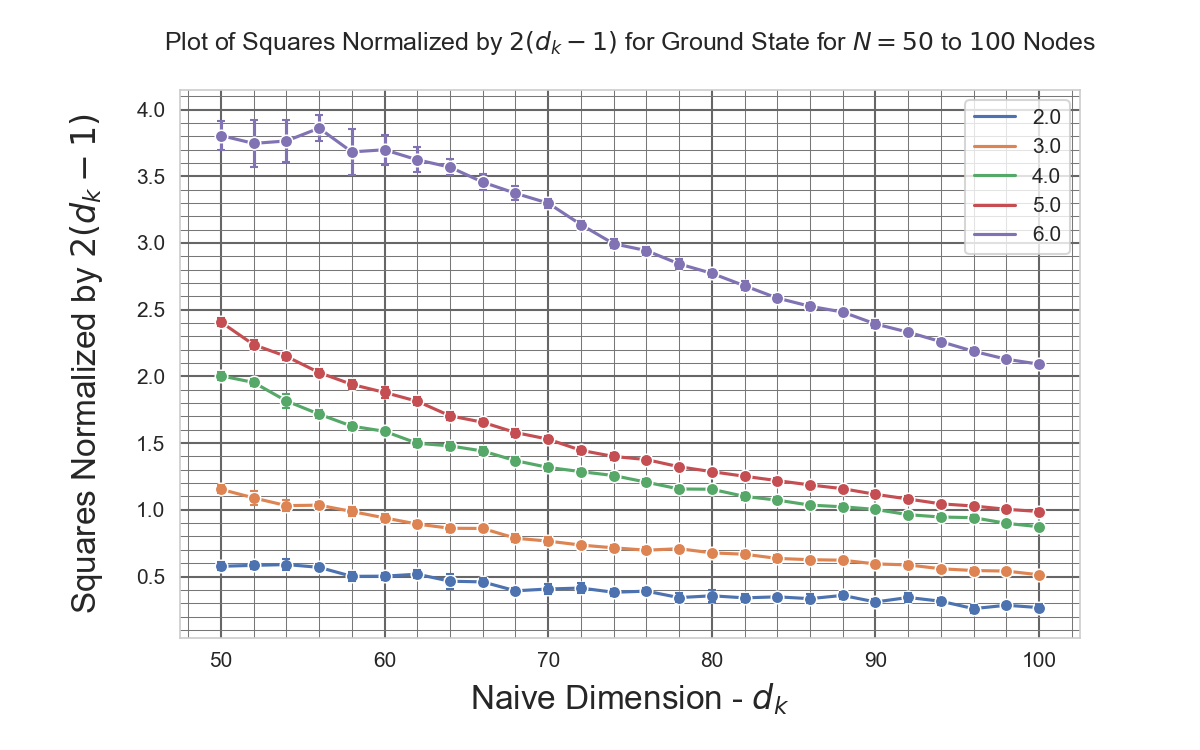}
		\caption{Normalized square cycle density versus naive dimension, for $H_{QMD}$ and $N=350$.}
		\label{fig:QMD-sqDens}
	\end{subfigure}
	~
	\begin{subfigure}[t]{0.45\textwidth}
		\centering
		\includegraphics[scale=0.38]{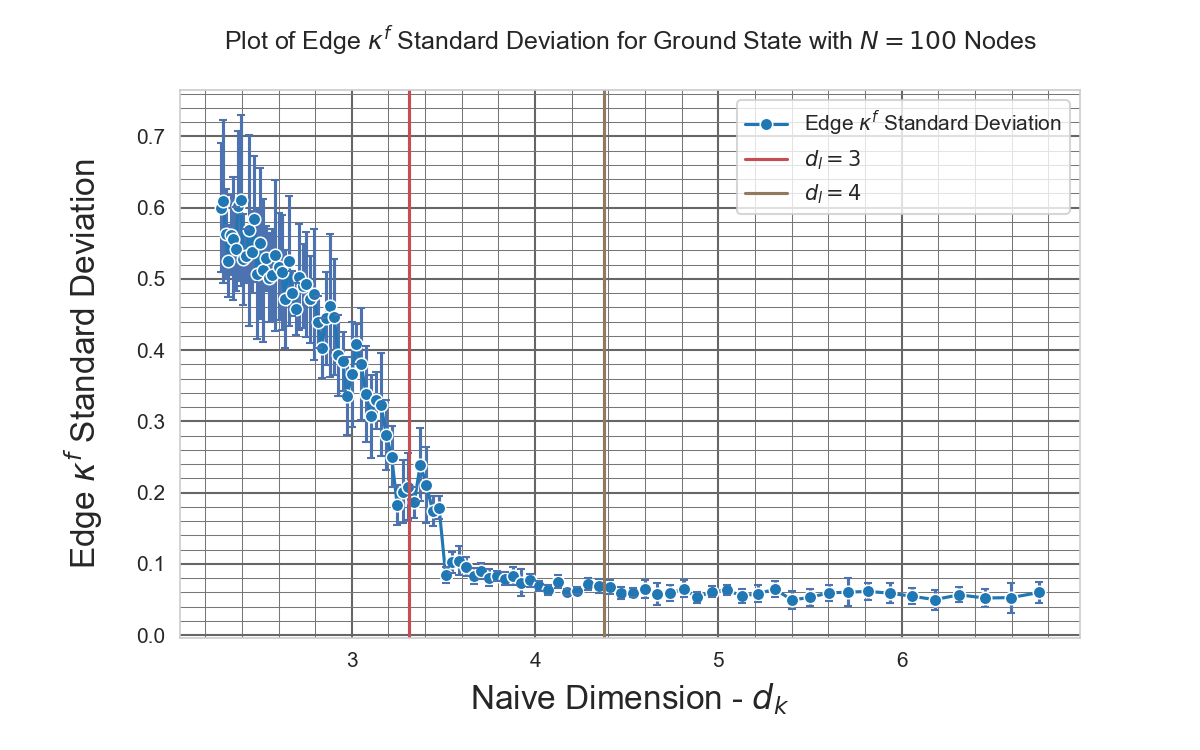}
		\caption{Standard deviation of Forman-Ricci curvature for $H_{CIM}$, $N=100$, as a fraction of the average edge curvature.}
		\label{fig:FR-kappaStd}	
	\end{subfigure}%
	~ 
	\begin{subfigure}[t]{0.45\textwidth}
		\centering
		\includegraphics[scale=0.38]{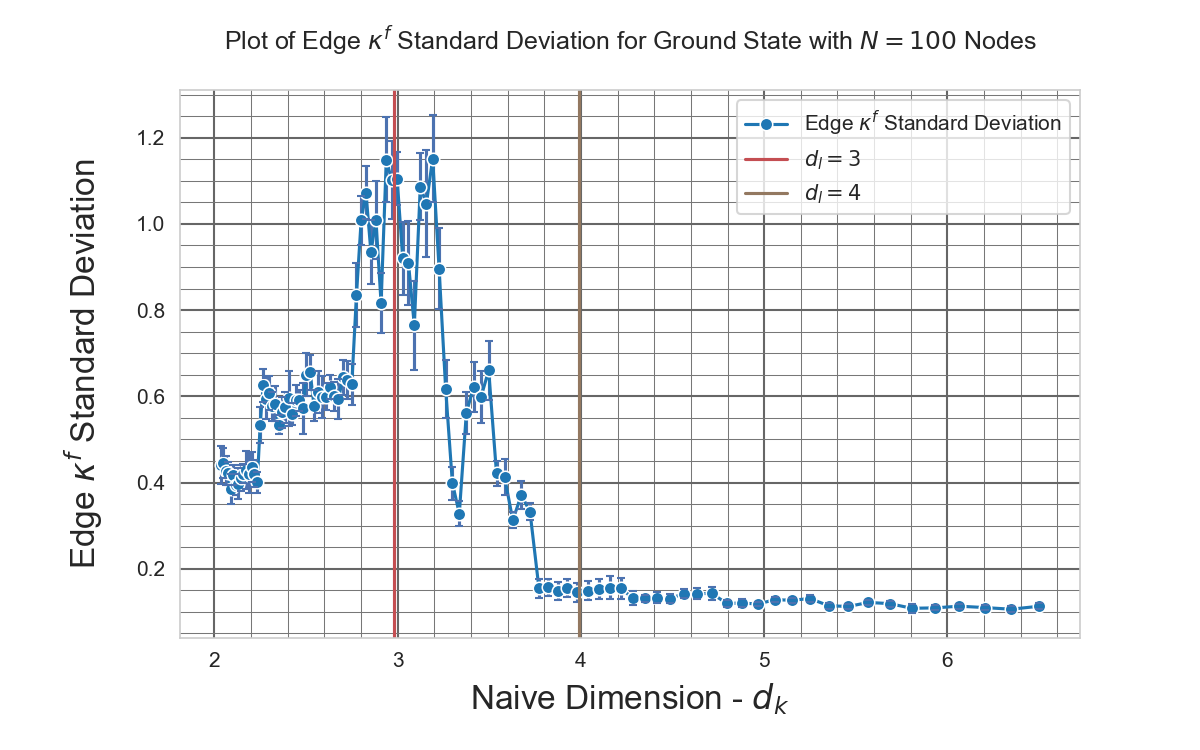}
		\caption{Standard deviation of Forman-Ricci curvature for $H_{QMD}$, $N=100$, as a fraction of the average edge curvature.}
		\label{fig:QMD-kappaStd}
	\end{subfigure}
	\caption{Comparison of key topology metrics for the ground states of both $H_{CIM}$ and $H_{QMD}$.  We plot all of these against the naive dimension \Cref{eqn:naive_dimension}. We compare the variation of dimensions, curvature regularity, polygon frequency and variation of polygon frequency with graph size. }
	\label{fig:ground_state_topo}
\end{figure*}

In order to probe the structure and topology of the ground states we present in \Cref{fig:ground_state_topo} a selection of metrics for both the CIM and QMD models.

In \Cref{fig:FR-Dims} and \Cref{fig:QMD-Dims} we plot the different measures of  dimension against the naive dimension $d_k$.
In both models the lattice dimension $d_l$ varies approximately linearly with $d_k$, although this linear relationship is much stronger with the QMD model.
This is to be expected, as the expression for $d_k$ in the CIM model is an approximation.
In both models there is a divergence of extrinsic and intrinsic measures of dimension as $d_k$ reduces below $4.5$.
This is consistent with earlier results \cite{trugenberger2015quantum,tee2020dynamics}, but the result for CIM is much more pronounced.
We have already remarked that for small values of $d_k$ the CIM model has a large amount of clustering and appears to produce a more chaotic ground state.
This more exaggerated divergence of dimensions in CIM is potential evidence of a `crumpled' phase of the ground state graph at low dimension, whereby instead of a smooth ground state a highly folded and non-local graph is obtained.
The non-locality will serve to increase the spectral dimension $d_S$ due to more rapid decay of return probability.
The folding will increase the Hausdorff dimension of the graph reflecting the higher dimension needed of a space in which the graph can be embedded.

Our model should ideally produce a $d_k$ dimensional square lattice $\mathbb{Z}^{d_k}$, in which each edge bounds $2(d_k-1)$ squares and pentagons and triangles are rare.
In \Cref{fig:FR-polygons} and \Cref{fig:QMD-polygons} we plot for a graph of $N=100$ nodes the average edge density of triangles, squares and pentagons.
For the squares we normalize by $2(d_k-1)$ and for visibility in \Cref{fig:QMD-polygons} the pentagon density is on a logarithmic scale.
In both cases squares dominate numerically in the graph, but critically between $d_l=3$ and $4$ the normalized value is approximately $1.0$.
A perfectly square lattice would have $\expval{\square_{ij}}/2(d_k-1) = 1.0$ identically, and we conclude that this regular geometry is only achieved at the same low dimension at which the intrinsic and extrinsic dimensions diverge.

The result is not exactly $1.0$ at either $d_l=3.0$ or $d_l=4.0$, but this could be because of finite size effects in the graph causing experimental error.
Accordingly in \Cref{fig:FR-sqDens} and \Cref{fig:QMD-sqDens} we plot the values of the edge square density for graphs of increasing size.
Across a range of naive dimensions from $d_k=2.0$ to $6.0$ we see for both models that $\expval{\square_{ij}}/2(d_k-1)$ reduces as the graph size increases.
It is difficult to form a conclusive opinion on the limiting behavior of graph size, but we remark that the value for $d_k=3.0$ appears to be stabilizing around $1.0$ and for QMD this occurs at $d_k=4.0$ and $5.0$.
Further investigations with much larger graphs are required to sharpen the result.

Finally, in \Cref{fig:FR-kappaStd} and \Cref{fig:QMD-kappaStd} we plot the variation with naive dimension of the standard deviation of the FR curvature on an edge, as a fraction of the the mean value.
We would expect, if the emergent discrete topology is a coarse grained version of a smooth manifold, the discrete topology of the graph to be highly uniform.
Consequently the curvature of the edges should also be highly uniform for a given value of $d_k$.
In the QMD model the standard deviation is low and stable at $d_l=4$, but before the dimension reduces to $d_l=3$ the variation becomes much more pronounced. 
In the case of CIM, the curvature is effectively uniform for $d_l>3$, but below this value is much less stable.
This result somewhat strengthens the hypothesis that our ground states are regular lattices, with the difference that the transition to a less regular and `crumpled' phase occurs at a lower dimension for CIM of around $d_l=3$.

We began our analysis with the goal of identifying whether the QMD and similar IG models are approximations of a more fundamental one, which in the low energy limit replicates the smooth spacetime of GR.
We believe the results presented here provide some support for this hypothesis, but with important caveats.
It is striking how similar the ground state properties of CIM and QMD are to each other. 
In most regards the connectivity, topology and regularity of the ground states of CIM and QMD are almost identical.
We may speculate, however, that whereas $d_l=4$ appears to be the dimension of QMD that has the high regularity and topology of a regular lattice, for CIM this is slightly lower and potentially at a value of  $d_l=3$.
If QMD is a coarse approximation of CIM, as indicated by our analysis resulting in \Cref{eqn:qmd_cim}, the refinements and extra terms in the Hamiltonian are responsible for this.
These originate directly from considerations of curvature, and have their origin in canonical GR.
Put another way, if our geometry is emergent and governed by a discrete form of GR at short scale, it would appear to prefer $3$ spatial dimensions.

The two major caveats are of course the non-rigorous nature of the analysis that took us from canonical GR to the CIM model, and the fact that the model is intrinsically non-relativistic.
Nowhere in our analysis have we factored in symmetry under the Lorentz group or demanded Lorentz covariance,  and we stress that our model is by definition not a real-world model of emergent geometry until this is remedied.
We note in passing that the interaction previously proposed as a model of dynamics in IG models \cite{tee2020dynamics}, exhibits an inherent maximum speed of propagation.
This maximum speed of propagation could be interpreted as an approximate form of causality \cite{tee2021quantum}, and may indicate that a relativistic version of CIM is within reach.

\section{Concluding remarks}
\label{sec:conclusion}
In this paper we have taken as our starting point the canonical formulation of GR and used the Hamiltonian in the Gaussian gauge to propose a discrete model of emergent geometry that we call the Canonical Ising model.
This model, like earlier Ising models of emergent geometry, describes a universe that is comprised of `atoms' of spacetime that can be thought of as qubits on both vertices and edges.
These edge and vertex states, as modeled in the Hamiltonian, interact to form stable ground states resembling a regular lattice.
When we numerically compute  the ground states of CIM, we find that they are highly regular and topologically similar to the QMD and similar models described.
We note however that in some ways CIM evolves with coupling in a smoother fashion than QMD.

The ground states are not spatially Forman-Ricci flat, but our freedom to choose a specific gauge or foliation in canonical GR, brings with it the Hamiltonian and momentum constraints.
This requires that $\mathcal{C}=~^3R +K^2-K^{ab}K_{ab}=0$, and so we see that negative spacial curvature of our hypersurfaces simply requires that $K^2-K^{ab}K_{ab} >0$.

What kind of universe does our model describe?
We remarked in \Cref{sec:results} that there is something special about $d_l=3$, which as our total manifold is $M=G(V,E) \bigotimes \mathbb{R}$ has the pleasing conclusion that  we are describing a $3+1$ dimensional total space, albeit with a Euclidean geometry.
However, the model also indicates that the higher energy regime has lower entropy and higher dimension.
As discussed in \Cref{sec:results}, it is tempting to interpret the results as we vary upwards the coupling constant from an initially small value, as representing the evolution of our model universe from a hot, low entropy, high connectivity and dimension initial state.
From this point the universe gradually cools and topologically `unfolds' into a nearly flat, and spatially three dimensional spatial geometry.
This mirrors our current understanding of the large scale geometry of the real Universe.
We speculate that this involves a topological `big bang', that progressed by the gradual reduction in the dimensionality of the universe, which can potentially provide an alternative interpretation of inflation and the cosmic microwave background uniformity.

In the initial state all points are local to all others, and by definition in causal contact.
As a consequence thermal equilibrium and uniformity are to be expected.
As dimensionality reduces, the size of the spatial extent of the graph, $r_U$, does not increase smoothly.
Due to the relationship of $r_U$ to the dimension, $r_U=V^{1/d}$, as $d$ reduces the extent of the universe accelerates.
This acceleration will initially be very slow for large $d$, but will accelerate as $d$ gets closer to $3$.
This expansion will continue whilst pockets of the graph have not reduced to $d=3$.
To summarize, we begin with a low entropy spatially `small' and energetically hot universe that is in thermal equilibrium that undergoes an accelerating increase in spatial extent as the dimensionality reduces with cooling.
If dimensional reduction comes to a halt as we approach $d=3$, this acceleration will slow down as more of the universe graph becomes uniform and of dimension $3$.
As all spacetime points are in causal contact prior to this topological `big bang' we would expect that all points in the model would be in equilibrium, and would expect in the evolved model for the energy of the spatial points to be uniform.
This is  a consistent story with the models of an inflationary universe, at least qualitatively.

The CIM model though is still not complete and is currently in our view a toy model of emergent geometry, albeit at least related to GR by virtue of its formulation.
Conspicuously it is non-relativistic, and describes an effectively Euclidean universe.
For it to be taken seriously this has to be remedied.
It is well know that discrete sub-groups of the the Lorentz group exist \cite{tarakanov2012some,foldes2008lorentz}, which have been speculated as providing a mechanism for a regular lattice to exhibit Lorentz symmetry, so it would seem feasible to construct one.
Alongside this Amelino-Camelia \cite{amelino2002relativity} has argued that a fundamental length scale can be reconciled with relativity in the `Doubly Special Relativity' model.
We believe that recasting the Ising structure of our model using Dirac spinors instead of qubits could provide a way of constructing a version of CIM that has Lorentz symmetry.
This is the subject of ongoing work.

What we believe we can safely conclude is that the well studied models of emergent geometry based upon Ising Hamiltonians may be approximate forms of a theory more closely related to general relativity, at least in a discretized form.
As the relationship between non-relativistic quantum mechanics and the dynamics of excitations in this model has already been shown in other work, this holds out the tantalizing prospect that GR itself is a low energy approximation of a quantum model of emergent geometry in which quantum mechanics is an integral feature.
If you admit the idea that the reconciliation of quantum mechanics and relativity requires one to be constructed from the other, perhaps this is a small hint that quantum mechanics could be used to emerge geometry and therefore general relativity.
A conclusive answer to this question is still far from being decided by the work presented here, but we believe that discrete emergent geometries will be an important part of the answer.

\clearpage
\bibliographystyle{JHEP}
\bibliography{CanonicalIsing-JHEP}

\end{document}